\documentclass[11pt,a4paper]{article}

\usepackage{jheppub}
\usepackage{amsmath,amsfonts,mathrsfs,amssymb}
\usepackage{epsfig}
\usepackage{latexsym}
\usepackage{pdfsync}
\usepackage{graphicx}
\usepackage{dcolumn}
\usepackage{bm}
\usepackage{bbm}
\usepackage[T1]{fontenc} 
\usepackage{epsf}
\usepackage{amsthm}
\usepackage{color}
\usepackage{slashed}
\usepackage{float}
\usepackage{esvect}
\usepackage{mathtools}
\usepackage{hyphenat}

\newbox\pippobox
\newcommand{\RNum}[1]{\uppercase\expandafter{\romannumeral #1\relax}}

\title{Spectra of glueballs and oddballs and the equation of state from holographic QCD}

\author[a]{Lin Zhang,}
\author[b]{Chutian Chen,}
\author[a]{Yidian Chen}
\author[a]{and Mei Huang}
\affiliation[a]{School of Nuclear Science and Technology, University of Chinese Academy of Sciences,\\Beijing, P.R.China 100049}
\affiliation[b]{School of Physical Sciences, University of Chinese Academy of Sciences,\\Beijing, P.R.China 100049}
\emailAdd{zhanglin@ucas.ac.cn}
\emailAdd{chenchutian18@mails.ucas.ac.cn}
\emailAdd{chenyidian@ucas.ac.cn}
\emailAdd{huangmei@ucas.ac.cn}

\abstract{We study the spectra of two-gluon glueballs and three-gluon oddballs and corresponding equation of state in $5$-dimensional deformed holographic QCD models in the graviton-dilaton system, where the metric, the dilaton field and dilaton potential are self-consistently solved from each other through the Einstein field equations and the equation of motion of the dilaton field. We compare the models by inputting the dilaton field, inputting the deformed metric and inputting the dilaton potential, and find that with only 2 parameters, the $5$-dimensional holographic QCD model predictions on glueballs/oddballs spectra in general are in good agreement with lattice results except two oddballs $0^{+-}$ and $2^{+-}$. From the results of glueballs/oddballs spectra at zero temperature and the equation of state at finite temperature, we observe that the model with quadratic dilaton field can simultaneously describe glueballs/oddballs spectra as well as equation of state of pure gluon system. The model with quadratic $A_{E}(z)$ can describe glueballs/oddballs spectra, but its corresponding equation of state behaves more like $N_{f}=2+1$ quark matter. These are consistent with dimension analysis at UV boundary.}


\begin{document}
\maketitle

\newpage

\section{Introduction}

Glueball is one of the most crucial predictions from quantum chromodynamics (QCD), whose non-Abelian feature makes it possible to form bound states of gauge bosons, i.e. glueballs made of two/three gluons (gg, ggg, etc.) \cite{Gell-Mann:2015noa}. The gauge field plays a more important dynamical role in glueballs than that in the standard hadrons, therefore studying particles like glueballs offers a good opportunity of understanding non\hyp{}perturbative aspects of QCD. The glueball spectra has attracted much attention for four decades \cite{Gell-Mann:2015noa}, and it has been widely investigated by using various non-perturbative methods. For example, glueballs have been studied by using lattice QCD \cite{Morningstar:1999rf,Lucini:2001ej,Meyer:2004gx,Chen:2005mg,Gregory:2012hu,Bennett:2017kga,Bennett:2020hqd,Athenodorou:2020ani,Bennett:2020qtj,Athenodorou:2021qvs}, by using effective models like flux tube model \cite{Isgur:1984bm} and MIT bag model \cite{Jaffe:1975fd,Barnes:1981kq,Barnes:1981kp,Carlson:1982er,Chanowitz:1982qj}, by using QCD sum rules \cite{Dominguez:1986td,Dominguez:1986zv,Latorre:1987wt,Narison:1988ts,Narison:1996fm,Narison:1997nw,Huang:1998wj,Narison:2008nj,Qiao:2014vva,Tang:2015twt,Pimikov:2016pag,Pimikov:2017xap,Qiao:2017jxc,Pimikov:2017bkk,Chen:2021cjr} as well as by using relativistic many\hyp{}body approach \cite{Szczepaniak:1995cw,Llanes-Estrada:2000ozq,Llanes-Estrada:2005bii}. There are also some other analyses of glueballs in Refs. \cite{Bugg:2000zy,Zhao:2005ip,Cheng:2006hu,Li:2009rk,He:2009sb,Cheng:2009zk,Janowski:2014ppa,Eshraim:2015cia,Sarantsev:2021ein}. For more information, please refer to review papers
\cite{Mathieu:2008me,Klempt:2007cp,Amsler:2004ps}.

On the other hand, the spin and mass of the glueball can be constrained from high energy scattering data. Regge trajectories $\alpha(t)=\alpha_0+\alpha't$ of the glueball have been used to fit high energy pp and $p \bar {p}$ scattering cross-section. The C-even glueball, Pomeron exchange gives the lightest $J=2^{++}$ glueball mass $M=\sqrt{t}=1.92~ GeV$. Analogy with the "Pomeron", C parity odd "Odderon" contributing to large odd amplitude was proposed in 1970s in describing the high energy $pp$ and $p{\bar p}$ scattering\cite{Braun:1998fs,Lukaszuk:1973nt}. The Odderon was regarded as three-gluon state:
\begin{equation}
  O_{abc}^{\mu\nu\sigma}(k1,k2,k3)=d_{abc}G_a^{\mu}(k_1)G_b^{\nu}(k2)G_c^{\nu}(k3)
\end{equation}
where the lower indices refer to color and the upper ones refer to the Lorentz structure, and $d_{abc}$ is the fundamental symmetric tensor in SU(3). The evidence for the identification of the odderon has been debated for a longtime. Recently, the D0 and TOTEM Collaborations announced the evidence of a t-channel exchanged C-odd odderons in $pp$ and $p{\bar p}$ scattering \cite{Abazov:2012qb,Abazov:2020rus}. Especially the odderon's contribution at the dip-bump region is very essential. The mass of $3^{--}$ odderon $M_{3^{--}}=3.001 \mathrm{GeV}$ and dacay width $\Gamma_{3^{--}}=2.984 \mathrm{GeV}$ are extracted by using the dipole (DP) Regge model to fit the scattering data \cite{Szanyi:2019kkn,Csorgo:2019ewn,Csorgo:2020wmw,Bence:2018ain}.

In Ref. \cite{Chen:2021cjr}, the oddball spectra has been calculated by using the QCD sum rule. In this work, we are going to investigate the glueball spectra in
the framework of holographic QCD, which is based on the gravity/gauge duality, or anti-de Sitter/conformal field theory (AdS/CFT) correspondence \cite{Maldacena:1997re,Gubser:1998bc,Witten:1998qj}.  AdS/CFT correspondence offers a new possibility to tackle the difficulty of strongly coupled gauge theories \cite{Aharony:1999ti,Aharony:2002up,Zaffaroni:2005ty,Erdmenger:2007cm}. Many efforts from both top-down and bottom-up approaches have been paid on examining the non\hyp{}perturbative properties of QCD \cite{Kovtun:2004de}, e.g.,  QCD equation of state, phase transitions, fluid properties of quark-gluon plasma, meson spectra \cite{Erlich:2005qh,Karch:2006pv,Sakai:2004cn,Sakai:2005yt,deTeramond:2005su,DaRold:2005mxj,Ghoroku:2005vt,Andreev:2007vn,Andreev:2006ct,Kruczenski:2004me,Kuperstein:2004yf,Forkel:2007cm}, baryon spectra \cite{Hong:2006ta,Nawa:2006gv,Hong:2007kx}, as well as the glueball sector \cite{Csaki:1998qr,deMelloKoch:1998vqw,Zyskin:1998tg,Minahan:1998tm,Csaki:1998cb,Csaki:1999uw,Brower:2000rp,BoschiFilho:2002vd,BoschiFilho:2002ta,Apreda:2003sy,BoschiFilho:2005yh,Colangelo:2007pt,Forkel:2007ru,FolcoCapossoli:2013eao,Bellantuono:2015fia,FolcoCapossoli:2015jnm,FolcoCapossoli:2016fzj,FolcoCapossoli:2016uns,Rodrigues:2016cdb,Rodrigues:2016kez,FolcoCapossoli:2019imm}. In Refs. \cite{Elander:2009bm,Elander:2010wd}, by linearizing the fluctuations around a classical $\sigma$-model coupled to gravity in $d+1$ dimensions, a gauge invariant (diffeomorphism invariant) formalism for calculating the spectra of scalar glueballs and tensor glueballs was developed, which was initially proposed in Refs. \cite{Bianchi:2003ug,Berg:2005pd,Berg:2006xy}. This algorithmic formalism was tested and some non-trivial applications were given in Refs. \cite{Elander:2012yh,Elander:2017cle,Elander:2017hyr,Elander:2018gte,Elander:2018aub,Elander:2020csd,Elander:2020ial,Elander:2020nyd,Elander:2020fmv,Elander:2021wkc}. The glueball mass spectra and decay rate in the Sakai-Sugimoto model have been investigated in Refs. \cite{Hashimoto:2007ze,Brunner:2015oqa,Brunner:2015yha}. Glueballs and oddballs spectra have also been widely studied by using the bottom-up approach, where some studies are based on hard-wall \cite{Erlich:2005qh} and soft-wall holographic QCD models \cite{Karch:2006pv} with the conformal $AdS_5$ background metric.

A realistic non-conformal holographic QCD model should reveal both the spontaneous chiral symmetry breaking and color charge confinement or linear confinement, which are two main features of QCD in the low energy regime. In the top-down approach, the Sakai-Sugimoto (SS) model or $D_4-D_8$ brane system \cite{Sakai:2004cn,Sakai:2005yt} is one of the most successful non-conformal holographic QCD models.
In the bottom-up approach, the dynamical holographic QCD (DhQCD) model constructed in Refs. \cite{Li:2012ay,Li:2013oda,Li:2013xpa} can simultaneously describe both chiral symmetry breaking and linear confinement, where the gluon dynamics background is solved by the coupling between the graviton and the dilaton field $\Phi(z)$, which is responsible for the gluon condensate and confinement, and the scalar field $X(z)$ is introduced to mimic chiral dynamics. Evolution of the dilaton field and scalar field in $5$-dimensional space\hyp{}time resemble the renormalization group from ultraviolet (UV) to infrared (IR). This dynamical holographic QCD model describes the scalar glueball spectra and the light meson spectral quite well \cite{Li:2012ay,Li:2013oda,Li:2013xpa}. Further studies \cite{Li:2014hja,Li:2014dsa,Chelabi:2015cwn} show that this dynamical holographic QCD model can also describe QCD phase transition, equation of state of QCD matter and temperature dependent transport properties, including shear viscosity, bulk viscosity, electric conductivity as well as jet quenching parameter. Except the dynamical holographic QCD model, there are several other non-conformal holographic QCD models in the same graviton-dilaton system which can well describe non-perturbative QCD propertities, e.g. the Gubser model \cite{Gubser:2008ny,Gubser:2008yx,DeWolfe:2010he} and the improved holographic QCD model \cite{Gursoy:2007cb,Gursoy:2007er,Gursoy:2010fj} with inputing of a dilaton potential, and the refined model \cite{Yang:2014bqa} and Dudal model \cite{Dudal:2017max} with inputting of a deformed metric.

In the graviton-dilaton system, the  metric, the dilaton field and the dilaton potential are self-consistently solved from each other through the Einstein field equations and the equation of motion of the dilaton field. In principle, the three types of models, A) inputting the form of the dilaton field, B) inputting the deformed metric, and C) inputting the dilaton potential, should be equivalent to describe the background at zero temperature and zero density. We will compare the glueball including (scalar, vector as well as tensor glueballs and their excitations) and oddball spectra, and compare thermodynamical properties with lattice QCD results for pure gluon system and/or $2+1$ flavors system in these three types of models.

The paper is organized as following: we introduce the general Einstein-Maxwell-dilaton framework in section \ref{sec-EMD}. Then in section \ref{five_models} we introduce five different models in the graviton-dilaton system. In section \ref{glueb_spect} we introduce the glueball and oddball operator and calculate the mass spectra in these models and we compare the results of mass spectra with lattice results, results from QCD sum rule and results extracted from high energy scattering data. In section \ref{eos} we compare thermodynamical properties of these models with lattice results. Finally, a short summary is given in section \ref{sec-sum}.

\section{The general Einstein-Maxwell-dilaton system}
\label{sec-EMD}

To keep the self-consistency of investigating the gluebal spectra as well as further studies on QCD matter at finite temperature and finite chemical potential, we firstly introduce the general framework of the Einstein-Maxwell-dilaton (EMD) system, which comes back to the graviton-dilaton coupling system at zero chemical potential. The total action of $5$-dimensional holographic QCD model including glueball/oddball excitations takes the following form:
\begin{align}
   & S_{\text {total }}^{s}=S_{b}^{s}+S_{g}^{s},
  \label{action_total}
\end{align}
where $S_{b}^{s}$ is the action for the background in the string frame, and $S_{g}^{s}$ is the action describing glueballs in the string frame.


The Einstein-Maxwell-dilaton action $S_{b}^{s}$ for the background in the string frame takes the form of:
\begin{alignat}{3}
   & S_{b}^{s}=\frac{1}{2 \kappa_{5}^{2}} \int \mathrm{d}^{5} x \sqrt{-g^{s}} e^{-2 \Phi} & \Big[ & R^{s}+4 {g^{s}}^{M N} \partial_{M} \Phi \partial_{N} \Phi-V^{s}(\Phi)
  \nonumber                                                                                                                                                                                                                                    \\
   &                                                                                      &       & -\frac{h(\Phi)}{4} e^{\frac{4 \Phi}{3}} {g^{s}}^{M \widetilde{M} }{g^{s}}^{N \widetilde{N} } F_{M N} F_{\widetilde{M} \widetilde{N}}\Big],
  \label{action_EMD_11}
\end{alignat}
where $s$ denotes the string frame, $\kappa_{5}^{2}=8 \pi G_5$, the $G_5$ is the $5$-dimensional Newton constant. The $g^s$ is the determinant of the metric in the string frame: $g^s=\det \left(g_{M N }\right)$, and the metric tensor in the string frame is extracted from
\begin{align}
  d s^{2}=\frac{L^2 e^{2 A_{s}(z)}}{z^{2}}\left(-f(z) d t^{2}+\frac{d z^{2}}{f(z)}+d y_{1}^{2}+d y_{3}^{2}+d y_{3}^{2}\right),
  \label{metric_str}
\end{align}
where $L$ is the curvature radius of the asymptotic $AdS_5$ space\hyp{}time. For simplicity, we set $L=1$ in the following calculations.
The $R^s$ is the Ricci curvature scalar in the string frame. The scalar field $\Phi(z)$ is the dilaton field which depends only on the coordinate $z$, $F_{M N}$ is the field strength of the $U(1)$ gauge field $A_{M}$:
\begin{align}
  F_{M N}=\partial_{M} A_{N}-\partial_{N} A_{M}.
  \label{field_strength_1}
\end{align}
The $5$-dimensional field $A_{M}$ is dual to baryon number current. $h(\Phi)$ describes the coupling strength of $A_{M}$ in the theory, $V^{s}(\Phi)$ represents the potential of the dilaton field in the string frame. $h(\Phi)$ and $V^{s}(\Phi)$ are the functions that depends only on the value of $\Phi$.

\subsection{The Einstein-Maxwell-dialton system in the Einstein frame}
As discussed in Ref. \cite{Li:2011hp},  it is convenient to calculate the vacuum expectation value of the loop operator in the string frame, and it is more convenient to work out the gravity solution and to study equation of state in the Einstein frame. So we apply the Weyl transformation \cite{weyl1921raum,weyl1993space}
\begin{align}
  g_{M N}^{s}=\mathrm{e}^{\frac{4}{3}\Phi}  g_{M N}^{E}
  \label{Weyl_transf}
\end{align}
to Eq. (\ref{action_EMD_11}). Here $g_{M N}^{E}$ is the metric tensor in the Einstein frame, the capital letter 'E' denotes the Einstein frame.
Then, Eq. (\ref{action_EMD_11}) can be written as
\begin{alignat}{3}
   & S^E =\frac{1}{2 \kappa_{5}^{2}} \int \mathrm{d}^{5} x \sqrt{-g^E} & \Big[ & R^E-\frac{4}{3} {g^E}^{M N} \partial_{M} \Phi \partial_{N} \Phi -V^E(\Phi)
    \nonumber                                                                                                                                                                                          \\
   &                                                                   &       & -\frac{h(\Phi)}{4} {g^{E}}^{M \widetilde{M} }{g^{E}}^{N \widetilde{N} } F_{M N} F_{\widetilde{M} \widetilde{N}}\Big],
  \label{action_EMD_2}
\end{alignat}
with $V^E=\mathrm{e}^{\frac{4}{3}\Phi} V^{s}$.

Then we define a new dilaton field $\phi$:
\begin{align}
   & \phi=\sqrt{\frac{8}{3}}\Phi.
  \label{new_dilaton}
\end{align}
Now Eq. (\ref{action_EMD_2}) becomes
\begin{alignat}{3}
   & S^E =\frac{1}{2 \kappa_{5}^{2}} \int \mathrm{d}^{5} x \sqrt{-g^E} & \Big[ & R^E-\frac{1}{2} {g^E}^{M N} \left( \partial_{M} \phi \right) \left( \partial_{N} \phi \right) -V_{\phi}(\phi)
  \nonumber                                                                                                                                                                                                   \\
   &                                                                   &       & -\frac{h_{\phi}(\phi)}{4} {g^{E}}^{M \widetilde{M} }{g^{E}}^{N \widetilde{N} } F_{M N} F_{\widetilde{M} \widetilde{N}}\Big],
  \label{action_EMD_3}
\end{alignat}
where $V_\phi(\phi)=V^E(\Phi)$, and $h_{\phi}(\phi)=h(\Phi)$. According to Eqs. (\ref{metric_str}), (\ref{Weyl_transf}) and (\ref{new_dilaton}), we can derive the line element in Einstein frame:
\begin{align}
  d s^{2}=\frac{L^2 e^{2 A_{E}(z)}}{z^{2}}\left(-f(z) d t^{2}+\frac{d z^{2}}{f(z)}+d y_{1}^{2}+d y_{3}^{2}+d y_{3}^{2}\right),
  \label{metric_Einstein}
\end{align}
where
\begin{align}
   & A_E(z)=A_s(z)-\sqrt{\frac{1}{6}}\phi(z).
  \label{relat_of_A}
\end{align}
After applying variation to Eq. (\ref{action_EMD_3}), we can derive the Einstein field equations and the equations of motion of $A_{M}$ and $\phi$ as follows
\begin{align}
   & R_{M N}^{E}-\frac{1}{2} g_{M N}^{E} R^{E}-T_{M N}=0, \nonumber                                                                                                                                    \\
   & \nabla_{M}\left[h_{\phi}(\phi) F^{M N}\right]=0, \nonumber                                                                                                                                        \\
   & \partial_{M}\left[\sqrt{-g} \partial^{M} \phi\right]-\sqrt{-g}\left(\frac{\mathrm{d} V_{\phi}(\phi)}{\mathrm{d} \phi}+\frac{F^{2}}{4} \frac{\mathrm{d} h_{\phi}(\phi)}{\mathrm{d} \phi}\right)=0,
  \label{EOMs_1}
\end{align}
with the energy-momentum tensor $T_{M N}$
\begin{alignat}{3}
   & T_{M N} & = & \frac{1}{2}\left[\left(\partial_{M} \phi \right) \left(\partial_{N} \phi \right) -\frac{1}{2} g_{M N}^{E} {g^E}^{P \widetilde{P}}\left(\partial_{P} \phi\right) \left(\partial_{\widetilde{P}} \phi\right)-g_{M N}^{E} V_{\phi}(\phi)\right]
  \nonumber                                                                                                                                                                                                                                                     \\
   &         &   & +\frac{h_{\phi}(\phi)}{2}\left({g^E}^{P \widetilde{P}} F_{M P} F_{N \widetilde{P}}-\frac{1}{4} g_{M N}^{E} {g^{E}}^{P \widetilde{P} }{g^{E}}^{Q \widetilde{Q} } F_{P Q} F_{\widetilde{P} \widetilde{Q}}\right).
  \label{energy_momentum_tensor_1}
\end{alignat}

We can safely suppose all the components of $A_{M}(z)$ are zero except $A_{t}(z)$. Substituting Eq. (\ref{metric_Einstein}) into the EOMs  Eq. (\ref{EOMs_1}), we then derive the EOMs for the components:
\begin{align}
   & A_{t}^{\prime \prime}+A_{t}^{\prime}\left(-\frac{1}{z}+\frac{{h_{\phi}}^{\prime}}{h_{\phi}}+{A_{E}}^{\prime}\right)=0,
  \label{EOMs_2_1}                                                                                                                                                                                                                                                                                     \\
   & f^{\prime \prime}+f^{\prime}\left(-\frac{3}{z}+3 {A_{E}}^{\prime}\right)-\frac{e^{-2 {A_{E}}} A_{t}^{\prime 2} z^{2} h_{\phi}}{L^2}=0,
  \label{EOMs_2_2}                                                                                                                                                                                                                                                                                     \\
   & A_{E}^{\prime \prime} +\frac{f^{\prime \prime}}{6 f}+A_{E}^{\prime}\left(-\frac{6}{z}+\frac{3 f^{\prime}}{2 f}\right)-\frac{1}{z}\left(-\frac{4}{z}+\frac{3 f^{\prime}}{2 f}\right)+3 {A_{E}}^{\prime 2} +\frac{L^2 e^{2 {A_{E}}} V_{\phi}}{3 z^{2} f}=0,
  \label{EOMs_2_3}                                                                                                                                                                                                                                                                                     \\
   & A_{E}^{\prime \prime}-A_{E}^{\prime}\left(-\frac{2}{z}+A_{E}^{\prime}\right)+\frac{\phi^{\prime 2}}{6}=0,
  \label{EOMs_2_4}                                                                                                                                                                                                                                                                                     \\
   & \phi^{\prime \prime}+\phi^{\prime}\left(-\frac{3}{z}+\frac{f^{\prime}}{f}+3 A_{E}^{\prime}\right)-\frac{L^2 e^{2 {A_{E}}}}{z^{2} f} \frac{\mathrm{d} V_{\phi}(\phi)}{\mathrm{d} \phi} +\frac{z^{2} e^{-2 {A_{E}}} A_{t}^{\prime 2}}{2 L^2 f} \frac{\mathrm{d} h_{\phi}(\phi)}{\mathrm{d} \phi}=0.
  \label{EOMs_2_5}
\end{align}
In the above 5 equations, only 4 of them are independent. Thus we can choose Eq. (\ref{EOMs_2_5}) as a constraint, which can be used to check the solutions.

\section{Five different models in the EMD system}
\label{five_models}

In the dilaton-graviton system, the metric, the dilaton field and the dilaton potential can be self-consistently solved from each other through the Einstein field equations and the equation of motion of the dilaton field.
At zero temperature and zero chemical potential, the function $f(z)=1$ and $A_{t}(z)=0$, then Eq. (\ref{EOMs_2_1}) to Eq. (\ref{EOMs_2_5}) can be simplified:
\begin{align}
   & A_{E}^{\prime \prime} -\frac{6}{z} A_{E}^{\prime}+\frac{4}{z^2}+3 {A_{E}}^{\prime 2} +\frac{L^2 e^{2 {A_{E}}} V_{\phi}}{3 z^{2}}=0,
  \label{EOMs_vac_1_1}                                                                                                                                                  \\
   & A_{E}^{\prime \prime}-A_{E}^{\prime}\left(-\frac{2}{z}+A_{E}^{\prime}\right)+\frac{\phi^{\prime 2}}{6}=0,
  \label{EOMs_vac_1_2}                                                                                                                                                  \\
   & \phi^{\prime \prime}+\phi^{\prime}\left(-\frac{3}{z}+3 A_{E}^{\prime}\right)-\frac{L^2 e^{2 {A_{E}}}}{z^{2}} \frac{\mathrm{d} V_{\phi}(\phi)}{\mathrm{d} \phi} =0,
  \label{EOMs_vac_1_3}
\end{align}
where Eq. (\ref{EOMs_vac_1_3}) is the constraint.
Under the condition that we have proper boundary conditions, if we input A) the form of the dilaton field $\phi(z)$, or B) the function $A_{E}(z)$, or C) the dilaton potential $V_{\phi}(\phi)$, we can solve the other two. In principle, these three types of models of EMD system are totally equivalent to describe the background in the vacuum. However, at finite temperature and finite chemical potential, the situation will become different. If we input $V_{\phi}(\phi)$, the form of $V_{\phi}(\phi)$ is independent of the temperature/chemical potential, from Eq. (\ref{EOMs_2_1}) and Eq. (\ref{EOMs_2_5}), we can solve different functions $A_{E}(z)$ and $\phi(z)$ at different temperature/chemical potential, which can be denoted as ${A_{E}}_{\,T,\, \mu}(T, \mu, z)$ and $\phi_{T,\, \mu}(T, \mu, z)$. On the other hand, if we input $A_{E}(z)$ (or $\phi(z)$), whose form is independent of temperature/chemical potential, we can derive $V_{\phi}(\phi)$ with temperature/chemical potential dependence, which can be denoted as ${V_{\phi}}_{\,T,\, \mu}(T, \mu, \phi)$. The two descriptions, that are equivalent at vacuum, now become distinct from each other at finite temperature/chemical potential. From now on, we call fixing $V_{\phi}(\phi)$  "description A", fixing $A_{E}(z)$ or $\phi(z)$ is denoted by "description B".

It is more convenient to solve the system in the Einstein frame from Eqs. (\ref{EOMs_vac_1_1}) $\sim$ (\ref{EOMs_vac_1_3}). In the following we list two sets of vacuum solutions of $V_{\phi}(\phi)$, $A_{E}(z)$ and $\phi(z)$ that satisfy the EOMs.

\subsection{Vacuum solutions: set \texorpdfstring{\RNum{1}}{1}}
\label{vac_sols_set_1}
From the experiences in Refs. \cite{Dudal:2017max,Yang:2014bqa}, we can input the function $A_E(z)$ in the Einstein frame, and solve $V_{\phi}(\phi)$ and $\phi(z)$. The simplest ansatz for the deformed metric is $A_{E}(z)=-a z^2$, and from Eqs. (\ref{EOMs_vac_1_1}) $\sim$ (\ref{EOMs_vac_1_3}) one can derive the solution as following:
\begin{align}
   & A_{E}(z)=-a z^2,
  \label{vev_sols_1_2}                                                                                                                    \\
   & V_{\phi}(\phi)=-\frac{6}{L^2} \mathrm{e}^{2 \left(k(\phi)\right)^2}\left(6 \left(k(\phi)\right)^4+5 \left(k(\phi)\right)^2+2\right),
  \label{vev_sols_1_1}                                                                                                                    \\
   & \phi(z)=z \sqrt{3 a\left(3+2 a z^{2}\right)}+\frac{3}{2} \sqrt{6} \operatorname{arcsinh} \left[\sqrt{\frac{2 a}{3}} z\right],
  \label{vev_sols_1_3}
\end{align}
where the auxiliary function $k(\varphi)$ is defined as the inverse function of
\begin{align}
   & \varphi(\mathfrak{z})={\mathfrak{z}} \sqrt{3 \left(3+2 {\mathfrak{z}}^{2}\right)}+\frac{3}{2} \sqrt{6} \operatorname{arcsinh} \left[\sqrt{\frac{2}{3}} {\mathfrak{z}}\right],
  \label{phi_to_scal_z}
\end{align}
which means $k(\varphi(\mathfrak{z}))=\mathfrak{z}$ with $\mathfrak{z}=\sqrt{a}z$. Starting from any of the above three functions, together with proper boundary conditions, we can solve other two functions from Eq. (\ref{EOMs_vac_1_1}) and Eq. (\ref{EOMs_vac_1_2}).

From Eq. (\ref{vev_sols_1_3}) we know that $\phi(z=0)=0$, $\lim_{z \to +\infty} \phi(z) \to +\infty$. At UV boundary ($z \to 0$), the asymptotic forms of $V_{\phi}(\phi)$ and $\phi$ are given below:
\begin{align}
   & L^2 V_{\phi}(\phi \to 0)=-12-\frac{3}{2}\phi ^2-\frac{1}{12}\phi ^4-\frac{377}{174960}\phi
  ^6-\frac{977}{33067440}\phi ^8-\frac{53483}{214277011200}\phi ^{10}
  \nonumber                                                                                         \\
   & \qquad \quad  -\frac{1564351}{1145524901875200}\phi ^{12}+\mathcal{O}\left({\phi}^{14}\right),
  \label{UV_asym_1_1}
\end{align}
\begin{align}
   & \phi(z \to 0)=6 \sqrt{a} z+\frac{2}{3} a^{\frac{3}{2}} z^3-\frac{1}{15} a^{\frac{5}{2}} z^5+\frac{1}{63} a^{\frac{7}{2}} z^7-\frac{5}{972} a^{\frac{9}{2}} z^9+\frac{7
  }{3564}a^{\frac{11}{2}}z^{11}
  \nonumber                                                                                                                                                                 \\
   & \qquad \quad -\frac{7 }{8424}a^{\frac{13}{2}} z^{13}+\mathcal{O}\left(z^{15}\right).
  \label{UV_asym_1_2}
\end{align}
From the UV asymptotic form of $V_{\phi}(\phi)$, we can extract the $5$-dimensional mass square of $\phi$
\begin{align}
  M_{\phi}^2=-3.
  \label{phi_5D_mass_1}
\end{align}
According to the mass-dimension relationship $M^2=(\Delta-p)(\Delta+p-4)$ and $p=0$, the dimension
\begin{align}
   & {\Delta_{\phi}}_{-}=1, \quad {\Delta_{\phi}}_{+}=3.
  \label{phi_dim_1}
\end{align}
At IR boundary ($z \to +\infty$), $V_{\phi}(\phi)$ and $\phi(z)$ behave as
\begin{align}
   & L^2 V_{\phi}(\phi \to +\infty)=-\frac{27}{4} \left(\frac{3}{8}\right)^{\frac{1}{4}} \mathrm{e}^{-\frac{3}{2}} \mathrm{e}^{\frac{\sqrt{6}}{3}\phi+\cdots} \left(\phi^{\frac{1}{2}}+\cdots\right),
  \label{IR_asym_1_1}
\end{align}
\begin{align}
   & \phi(z \to +\infty)=\sqrt{6}\bigg[a z^2+\frac{3}{4}\left(1+\ln\left(\frac{8}{3}\right)+\ln\left(a z^2\right) \right)+\frac{9}{32 }\frac{1}{a z^2}-\frac{27}{256 }\frac{1}{a^2 z^4}+\frac{135}{2048 }\frac{1}{a^3 z^6}
    \nonumber                                                                                                                                                                                                              \\
   & \qquad \qquad \qquad -\frac{1701}{32768 }\frac{1}{a^4 z^8}+\frac{15309}{327680 }\frac{1}{a^5 z^{10}}-\frac{24057}{524288 }\frac{1}{a^6 z^{12}}+\mathcal{O}\left(\frac{1}{ z^{14}}\right)\bigg].
  \label{IR_asym_1_2}
\end{align}
Eq. (\ref{vev_sols_1_1}) lead to the masses of glueballs $m_{n}$ behave as
\begin{align}
  {m}_{n} \sim {n}^{\frac{1}{2}}, \qquad \text{when} \quad n \to +\infty.
  \label{glueb_mass_1}
\end{align}
which shows the linear Regge behavior along $n$.

\subsection{Vacuum solutions: set \texorpdfstring{\RNum{2}}{2}}
\label{vac_sols_set_2}
As for another set of solution, we start from the form of $\phi(z)$. One simple but nontrivial ansatz is to take the quadratic form of $\phi(z)$: $\phi(z)=b z^2$. As discussed in Refs. \cite{Li:2011hp,Li:2012ay,Li:2013oda,Li:2013xpa}, the quadratic form of the dilaton field is dual to a dimension-2 gluon condensation operator, which is responsible for the linear confinement of the gluon system.  Then the solution $V_{\phi}(\phi)$, $A_{E}(z)$ and $\phi(z)$ take the form of
\begin{align}
   & \phi(z)=b z^2,
  \label{vev_sols_2_3}                                                                                                                                                                                                                                                                                    \\
   & V_{\phi}(\phi)=\frac{1}{L^2}2\times 2^{\frac{3}{4}} \times 3^{\frac{1}{4}} \phi ^{\frac{3}{2}} \left[\Gamma \left(\frac{5}{4}\right)\right]^2 \left\{\left[I_{\frac{1}{4}}\left(\frac{\phi }{\sqrt{6}}\right)\right]^2-4\left[I_{-\frac{3}{4}}\left(\frac{\phi }{\sqrt{6}}\right)\right]^2 \right\},
  \label{vev_sols_2_1}                                                                                                                                                                                                                                                                                    \\
   & A_{E}(z)=-\ln \left[\frac{2^{\frac{3}{8}} \times 3^{\frac{1}{8}} \Gamma \left(\frac{5}{4}\right) I_{\frac{1}{4}}\left(\frac{b z^2}{\sqrt{6}}\right)}{{b^{\frac{1}{4}}
          \sqrt{z}}}\right],
  \label{vev_sols_2_2}
\end{align}
where $\Gamma(z)$ is the Euler gamma function, $I_{n}(z)$ is the modified Bessel function of the first kind.

From Eq. (\ref{vev_sols_2_3}) we know that $\phi(z=0)=0$, $\lim_{z \to +\infty} \phi(z) \to +\infty$. At UV boundary ($z \to 0$), the asymptotic form are
\begin{align}
   & L^2 V_{\phi}(\phi \to 0)=-12-2 {\phi}^2-\frac{4 }{15} {\phi} ^4-\frac{49
  }{6075}{ \phi} ^6-\frac{11 }{94770}{ \phi} ^8-\frac{11 }{11153700}{ \phi} ^{10}
  \nonumber                                                                                                 \\
   & \qquad \qquad \qquad \quad -\frac{38 }{6851160225}{ \phi} ^{12}+\mathcal{O}\left({ \phi} ^{14}\right),
  \label{UV_asym_2_1}
\end{align}
\begin{align}
   & A_{E}(z \to 0)=-\frac{1 }{30}b^2 z^4+\frac{1 }{4050}b^4 z^8-\frac{4  }{1184625}b^6 z^{12}+\mathcal{O}\left(z^{16}\right).
  \label{UV_asym_2_2}
\end{align}
From the UV asymptotic form of $V_{\phi}(\phi)$, we can extract the $5$-dimensional mass square of $\phi$
\begin{align}
  M_{\phi}^2=-4.
  \label{phi_5D_mass_2}
\end{align}
According to the mass-dimension relationship $M^2=(\Delta-p)(\Delta+p-4)$ and $p=0$, the dimension
\begin{align}
   & {\Delta_{\phi}}_{-}= {\Delta_{\phi}}_{+}=2.
  \label{phi_dim_2}
\end{align}
At IR boundary ($z \to +\infty$), $V_{\phi}(\phi)$ and $A_{E}(z)$ behave as
\begin{align}
   & L^2 V_{\phi}(\phi \to \infty)=-\frac{1}{\pi} {2}^{\frac{5}{4}} \times {3}^{\frac{7}{4}} \left[\Gamma \left(\frac{5}{4}\right)\right]^2 \mathrm{e}^{\frac{\sqrt{6}}{3}\phi} \phi ^{\frac{1}{2}} \bigg[1-\frac{23 \sqrt{6}}{48}\frac{1}{\phi}-\frac{277}{256}\frac{1}{{\phi}^{2}}-\frac{4365\sqrt{6}}{4096}\frac{1}{{\phi}^{3}}
    \nonumber                                                                                                                                                                                                                                                                                                                      \\
   & \qquad \qquad \quad -\frac{1271565}{131072}\frac{1}{{\phi}^{4}}-\frac{41182155 \sqrt{6}}{2097152}\frac{1}{{\phi}^{5}}-\frac{9973379745}{33554432}\frac{1}{{\phi}^{6}}-\frac{481731815565 \sqrt{6}}{536870912}\frac{1}{{\phi}^{7}}
    \nonumber                                                                                                                                                                                                                                                                                                                      \\
   & \qquad \qquad \quad -\frac{650211981544125}{34359738368}\frac{1}{{\phi}^{8}}-\frac{41734532955290175 \sqrt{6}}{549755813888}\frac{1}{{\phi}^{9}}
    \nonumber                                                                                                                                                                                                                                                                                                                      \\
   & \qquad \qquad \quad -\frac{18065483595471987675}{8796093022208}\frac{1}{{\phi}^{10}}-\frac{1447278481564158318075 \sqrt{6}}{140737488355328}\frac{1}{{\phi}^{11}}
    \nonumber                                                                                                                                                                                                                                                                                                                      \\
   & \qquad \qquad \quad -\frac{1529878341359963470300425}{4503599627370496}\frac{1}{{\phi}^{12}}-\frac{146978223450520872139104375 \sqrt{6}}{72057594037927936}\frac{1}{{\phi}^{13}}
    \nonumber                                                                                                                                                                                                                                                                                                                      \\
   & \qquad \qquad \quad +\mathcal{O}\left(\frac{1}{{\phi}^{14}}\right)\bigg] \left\{1+\frac{5 \sqrt{2}}{6} {\mathrm{e}}^{-\frac{\sqrt{6}}{3} \phi}\left[1+\mathcal{O}\left(\frac{1}{\phi}\right) \right] \right\},
  \label{IR_asym_2_1}
\end{align}
\begin{align}
   & A_{E}(z \to \infty)=-\frac{\sqrt{6}}{6} b z^2+\frac{3}{2}\ln\left(\sqrt{b}z\right)+\ln \left[\frac{{\pi}^{\frac{1}{2}}}{2^{\frac{1}{8}} 3^{\frac{3}{8}} \Gamma \left(\frac{5}{4}\right)}\right]-\frac{3 \sqrt{6}}{32} \frac{1}{b z^2}-\frac{9}{32}\frac{1}{b^2 z^4}
  \nonumber                                                                                                                                                                                                                                                              \\
   & \qquad \qquad \quad -\frac{297 \sqrt{6}}{1024}\frac{1}{b^3 z^6}-\frac{1377}{512}\frac{1}{b^4 z^8}-\frac{451251 \sqrt{6}}{81920}\frac{1}{b^5 z^{10}}-\frac{172287}{2048}\frac{1}{b^6 z^{12}}+\mathcal{O}\left(\frac{1}{z^{14}}\right)
  \nonumber                                                                                                                                                                                                                                                              \\
   & \qquad \qquad \quad +\mathcal{O}\left({\mathrm{e}}^{-\frac{\sqrt{6}}{3} b z^2}\right)+\mathcal{O}\left(\frac{1}{z^2} {\mathrm{e}}^{-\frac{\sqrt{6}}{3} b z^2}\right).
  \label{IR_asym_2_2}
\end{align}
Again, from the asymptotic expansion of $V_{\phi}(\phi)$ at IR boundary, we can conclude that linear Regge behavior of the masses of glueballs $m_{n}$:
\begin{align}
  {m}_{n}^2 \sim n, \qquad \text{when} \qquad n \to +\infty,
  \label{glueb_mass_2}
\end{align}
which shows the linear Regge behavior along $n$.

\subsection{Five different models}

The two sets of vacuum solutions listed above have linear confinement and can produce glueball bound state. Not all models can show such feature. According to Refs. \cite{Gursoy:2010fj,Gursoy:2008za}, if we require that the theory is confined and bad singularities are absent, the asymptotic behavior of $V_{\phi}(\phi)$ at IR boundary should be
\begin{align}
   & L^2 V_{\phi}(\phi \to +\infty)=c_{V}\, \mathrm{e}^{c_{\phi,1}\phi+\cdots} \left(\phi^{c_{\phi,2}}+\cdots\right),
  \nonumber                                                                                                           \\
   & \left\{
  \begin{aligned}
     & \frac{\sqrt{6}}{3} < c_{\phi,1} < \frac{2\sqrt{3}}{3}, \quad c_{\phi,2} \,\,\text{is real number},
    \\
     & c_{\phi,1}=\frac{\sqrt{6}}{3}, \quad c_{\phi,2} \geqslant 0,
  \end{aligned}
  \right.
  \label{confinement_and_absence_of_bad_singularities_bound}
\end{align}
where $c_{V}$ is  constant. When $c_{\phi,1}=\frac{\sqrt{6}}{3}$ and $c_{\phi,2} > 0$, there are asymptotically linear glueball spectra:
\begin{align}
  m_{n} \sim n^{c_{\phi,2}} \qquad \text{when} \quad n \to +\infty.
  \label{asymp_linear_glueb_spect}
\end{align}

\begin{figure}[htb]
  \begin{center}
    \includegraphics[width=120mm,clip=true,keepaspectratio=true]{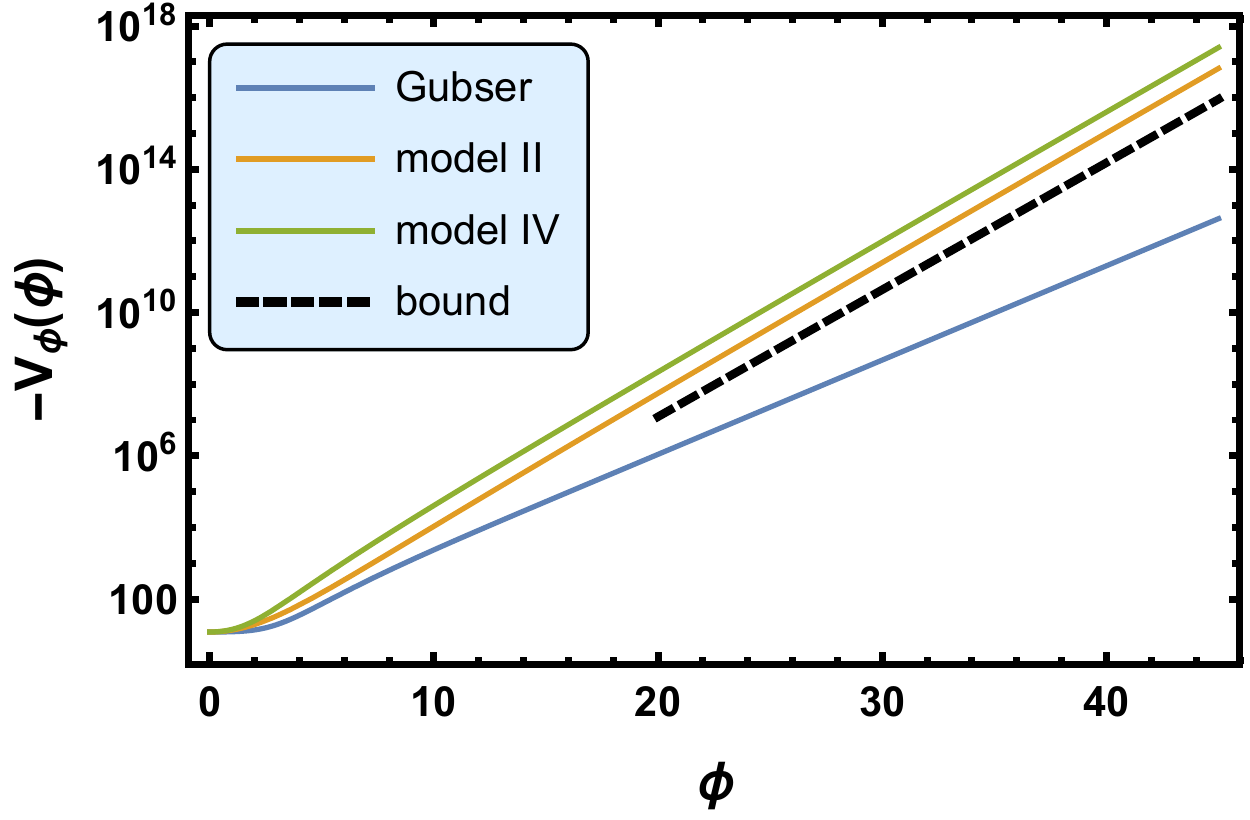}
  \end{center}
  \caption{These are three different dilaton potentials $V_{\phi}(\phi)$. The longitudinal axis is the value of $-V_{\phi}(\phi)$ in logarithm coordinate. The horizontal axis is the value of $\phi$. The dashed black line is ${\mathrm{e}}^{\frac{\sqrt{6}}{3} \phi}$. The blue line, orange line and green line represent the potential in Eq. (\ref{dilaton_pot_Gubser}), Eq. (\ref{vev_sols_1_1}) and Eq. (\ref{vev_sols_2_1}) respectively. The meaning of "model \RNum{2}" and "model \RNum{4}" will be explained later. The dashed black line is ${\mathrm{e}}^{\frac{\sqrt{6}}{3} \phi}$. The bound is given by Eq.(\ref{asymp_linear_glueb_spect}) from Refs. \cite{Gursoy:2010fj,Gursoy:2008za}. If the potential is more gradual than this bound, the theory is gapless and non-confining.}
  \label{dilaton_potentials_plot}
\end{figure}

For comparison, we plot three different dilaton potentials $V_{\phi}(\phi)$ in Fig. \ref{dilaton_potentials_plot}. One of them is the Gubser model taken from Ref. \cite{DeWolfe:2010he}:
\begin{align}
  V_{\phi}(\phi)=\frac{-12 \cosh{\left( 0.606\phi\right)}+2.057 {\phi}^{2}}{L^2},
  \label{dilaton_pot_Gubser}
\end{align}
the others two are Eq. (\ref{vev_sols_1_1}) and Eq. (\ref{vev_sols_2_1}). Here we set $L=1$. The dashed black line is ${\mathrm{e}}^{\frac{\sqrt{6}}{3} \phi}$. According to the conclusion in subsection \ref{vac_sols_set_1}, if the potential is more gradual than this line, such as the blue line that represents the Gubser model in Eq. (\ref{dilaton_pot_Gubser}), the theory is gapless and non-confining.

As we stated below Eqs. (\ref{EOMs_vac_1_1}) $\sim$ (\ref{EOMs_vac_1_3}), there are two different descriptions of the input of EMD system. Combining with the two different sets of vacuum solutions Eqs. (\ref{vev_sols_1_1}) $\sim$ (\ref{vev_sols_1_3})  and Eq. (\ref{vev_sols_2_1}) $\sim$ (\ref{vev_sols_2_3}), we consider five models in this article.

\subsubsection{Model \texorpdfstring{\RNum{1}}{1} and \texorpdfstring{\RNum{2}}{2}}
In model \RNum{1}, we use description-B and input $A_E(z)$ as Eq. (\ref{vev_sols_1_2}):
\begin{align}
   & A_{E}(z)=-a z^2.
  \label{input_model_1}
\end{align}
Note that the dimension of the parameter $a$ is ${\left[E\right]}^{2}$ and its value decides the energy scale of the EMD system.
At vacuum, the boundary condition is chosen as
\begin{align}
   & \phi(z=0)=0.
  \label{bound_cond_vac_model_1}
\end{align}
Combining the boundary condition Eq. (\ref{bound_cond_vac_model_1}) with the EOMs Eq. (\ref{EOMs_vac_1_2}) and Eq. (\ref{EOMs_vac_1_1}), we can solve $\phi(z)$ and $V_{\phi}(\phi)$. The results are Eq. (\ref{vev_sols_1_3}) and Eq. (\ref{vev_sols_1_1}).

As for finite temperature and finite chemical potential, the EOMs are Eq. (\ref{EOMs_2_1}) $\sim$ (\ref{EOMs_2_4}). There may exist the black hole
\footnote{ Actually, the black hole may not exist for all values of temperature and chemical potential. Under some situations, it only exists above a certain temperature $T_{\text{min}}$. The $AdS$ thermal gas is the only solution below $T_{\text{min}}$. However, under other situations, we always have the black hole solution.
}
in space\hyp{}time manifold, the metric of which in conformal coordinate $z$ is Eq. (\ref{metric_Einstein}). The boundary conditions are given as
\begin{align}
   & A_t(z=0)=\mu,
  \nonumber        \\
   & A_t(z=z_h)=0,
  \nonumber        \\
   & f(z=0)=1,
  \nonumber        \\
   & f(z=z_h)=0,
  \nonumber        \\
   & \phi(z=0)=0,
  \label{bound_cond_BH_model_1}
\end{align}
where $z=z_h$ is the location of the event horizon of black hole on the coordinate $z$, $\mu$ is the chemical potential. Besides the boundary condition Eq. (\ref{bound_cond_BH_model_1}), the form of $h_{\phi}(\phi)$ are also needed to solve the EOMs. However, we consider the $\mu=0$ case, which means $A_t(z)=0$ through the article. Thus our calculations and results are independent on $h_{\phi}(\phi)$.

\label{model_2}
In model \RNum{2}, we use description-A and  input $V_{\phi}(\phi)$ as Eq. (\ref{vev_sols_1_1}):
\begin{align}
   & V_{\phi}(\phi)=-\frac{6}{L^2} \mathrm{e}^{2 \left(k(\phi)\right)^2}\left(6 \left(k(\phi)\right)^4+5 \left(k(\phi)\right)^2+2\right).
  \label{input_model_2}
\end{align}
At vacuum, the boundary conditions are chosen as
\begin{align}
   & A_E(z=0)=0,
  \label{bound_cond_vac_model_2_index_1}                                  \\
   & \phi(z=0)=0,
  \label{bound_cond_vac_model_2_index_2}                                  \\
   & \left. \frac{\mathrm{d}\phi(z)}{\mathrm{d}z}\right|_{z=0}=6\sqrt{a}.
  \label{bound_cond_vac_model_2_index_3}
\end{align}
Eq. (\ref{bound_cond_vac_model_2_index_1}) guarantees the space\hyp{}time is asymptotic $AdS_{5}$ at UV boundary. Eq. (\ref{bound_cond_vac_model_2_index_3}) contains a parameter $a$, the dimension of which is ${\left[E\right]}^{2}$ and the value of which decides the energy scale of the EMD system. Given by these boundary conditions and set the value of $a$ in Eq. (\ref{bound_cond_vac_model_2_index_3}) to be the same with that in Eq. (\ref{input_model_1}) , we can solve the EOMs at vacuum, then it will be found that the solutions are totally equivalent to those in model \RNum{1} at vacuum.

As for finite temperature and finite chemical potential, the EOMs are Eq. (\ref{EOMs_2_1}) $\sim$ (\ref{EOMs_2_4}). The boundary conditions are given as
\begin{align}
   & A_t(z=0)=\mu,
  \nonumber                                                               \\
   & A_t(z=z_h)=0,
  \nonumber                                                               \\
   & f(z=0)=1,
  \nonumber                                                               \\
   & f(z=z_h)=0,
  \nonumber                                                               \\
   & A_E(z=0)=0,
  \nonumber                                                               \\
   & \phi(z=0)=0,
  \nonumber                                                               \\
   & \left. \frac{\mathrm{d}\phi(z)}{\mathrm{d}z}\right|_{z=0}=6\sqrt{a},
  \label{bound_cond_BH_model_2}
\end{align}
where $z=z_h$ is the location of the event horizon of black hole on the coordinate $z$, $\mu$ is the chemical potential. We should emphasize here that at finite temperature or finite chemical potential case, the solutions here are different from those in model \RNum{1}.

\subsubsection{Model \texorpdfstring{\RNum{3}}{3} and \texorpdfstring{\RNum{4}}{4}}
In model \RNum{3}, we use description B and  input $\phi(z)$ as Eq. (\ref{vev_sols_2_3}):
\begin{align}
   & \phi(z)=b z^2.
  \label{input_model_3}
\end{align}
Note that the dimension of the parameter $b$ is ${\left[E\right]}^{2}$ and its value decides the energy scale of the EMD system.
Substituting Eq. (\ref{input_model_3}) into Eq. (\ref{EOMs_vac_1_2}), we can solve a general solution for $A_E(z)$ with two integration constants.
However, the value of this general solution is usually complex. If we force the reality of $A_E(z)$ and consider the boundary condition
\begin{align}
   & A_E(z=0)=0,
  \label{bound_cond_vac_model_3}
\end{align}
we can solve $A_E(z)$ and $V_{\phi}(\phi)$. The results are Eq. (\ref{vev_sols_2_2}) and Eq. (\ref{vev_sols_2_3}).

As for finite temperature and finite chemical potential, the EOMs are Eqs. (\ref{EOMs_2_1}) $\sim$ (\ref{EOMs_2_4}). The boundary conditions are imposed as
\begin{align}
   & A_t(z=0)=\mu,
  \nonumber        \\
   & A_t(z=z_h)=0,
  \nonumber        \\
   & f(z=0)=1,
  \nonumber        \\
   & f(z=z_h)=0,
  \label{bound_cond_BH_model_3}
\end{align}
where $z=z_h$ is the location of the event horizon of black hole on the coordinate $z$, $\mu$ is the chemical potential. Collecting these boundary conditions and Eq. (\ref{input_model_3}), Eq. (\ref{vev_sols_2_2}), we can then solve the EMD system.

\label{model_4}
In model \RNum{4}, we use description A and input $V_{\phi}(\phi)$ as Eq. (\ref{vev_sols_2_1}):
\begin{align}
   & V_{\phi}(\phi)=\frac{1}{L^2}2\times 2^{\frac{3}{4}} \times 3^{\frac{1}{4}} \phi ^{\frac{3}{2}} \left[\Gamma \left(\frac{5}{4}\right)\right]^2 \left\{\left[I_{\frac{1}{4}}\left(\frac{\phi }{\sqrt{6}}\right)\right]^2-4\left[I_{-\frac{3}{4}}\left(\frac{\phi }{\sqrt{6}}\right)\right]^2 \right\}.
  \label{input_model_4}
\end{align}
At vacuum, the boundary conditions are chosen as
\begin{align}
   & A_E(z=0)=0,
  \label{bound_cond_vac_model_4_index_1}  \\
   & \lim_{z \to 0}\frac{\phi(z)}{z^2}=b,
  \label{bound_cond_vac_model_4_index_2}
\end{align}
Again, we force that $A_E(z)$ is real. The Eq. (\ref{bound_cond_vac_model_4_index_1}) guarantees the space\hyp{}time is asymptotic $AdS_{5}$ at UV boundary. Eq. (\ref{bound_cond_vac_model_4_index_2}) contains a parameter $b$, the dimension of which is ${\left[E\right]}^{2}$ and the value of which decides the energy scale of the EMD system. Given by these boundary conditions and set the value of $b$ in Eq. (\ref{bound_cond_vac_model_4_index_2}) to be the same with that in Eq. (\ref{input_model_3}), we can solve the EOMs at vacuum, then it will be found that the solutions are totally equivalent to those in model \RNum{3} at vacuum.

As for finite temperature and finite chemical potential, the EOMs are Eqs. (\ref{EOMs_2_1}) $\sim$ (\ref{EOMs_2_4}). The boundary conditions are given as
\begin{align}
   & A_t(z=0)=\mu,
  \nonumber        \\
   & A_t(z=z_h)=0,
  \nonumber        \\
   & f(z=0)=1,
  \nonumber        \\
   & f(z=z_h)=0,
  \nonumber        \\
   & A_E(z=0)=0,
  \nonumber        \\
   & \phi(z=0)=0.
  \label{bound_cond_BH_model_4}
\end{align}
Given by these boundary conditions, we still have the freedom to choose the energy scale of the EMD system. We should emphasis here that at finite temperature or finite chemical potential case, the solutions here are different from those in model \RNum{3}.

\subsubsection{Model \texorpdfstring{\RNum{5}}{5}}
In model \RNum{5}, we  input $\phi(z)$ as
\begin{align}
   & \phi(z)=\frac{2 \sqrt{6}}{3} z \sqrt{3 d\left(3+2 d z^{2}\right)}+ 6 \operatorname{arcsinh} \left[\sqrt{\frac{2 d}{3}} z\right].
  \label{input_model_5}
\end{align}
Note that the dimension of the parameter $d$ is ${\left[E\right]}^{2}$ and its value decides the energy scale of the EMD system.
Substituting Eq. (\ref{input_model_5}) into Eq. (\ref{EOMs_vac_1_2}), we force the reality of $A_E(z)$ and consider the boundary condition
\begin{align}
   & A_E(z=0)=0,
  \label{bound_cond_vac_model_5}
\end{align}
we can solve $A_E(z)$ and $V_{\phi}(\phi)$ numerically.
Although we can't get the analytical form of $A_E(z)$, we can still can derive its asymptotic expansions:

\begin{align}
   & A_{E}(z \to 0)=-\frac{8}{3}d z^2+\frac{8 }{9}d^2 z^4-\frac{512}{567} d^3 z^6+\frac{1664}{1701}d^4 z^8-\frac{311296}{280665}d^5 z^{10}
  \nonumber                                                                                                                                \\
   & \qquad \qquad \qquad +\frac{19972096}{15324309}d^6 z^{12}+\mathcal{O}\left( z^{14}\right),
  \label{AE_UV_asym_model_5}
\end{align}
As for finite temperature and finite chemical potential, the EOMs are Eqs. (\ref{EOMs_2_1}) $\sim$ (\ref{EOMs_2_4}). The boundary conditions are imposed as
\begin{align}
   & A_t(z=0)=\mu,  \nonumber \\
   & A_t(z=z_h)=0,  \nonumber \\
   & f(z=0)=1,  \nonumber     \\
   & f(z=z_h)=0,
  \label{bound_cond_BH_model_5}
\end{align}
where $z=z_h$ is the location of the event horizon of black hole on the coordinate $z$, $\mu$ is the chemical potential. Collecting these boundary conditions, the Eq. (\ref{input_model_5}), and the numerical solution of $A_{E}(z)$, we can then solve the EMD system.

\section{Spectra of glueballs and oddballs}
\label{glueb_spect}

In this section, we discuss the spectra of glueballs and oddballs. In our treatment, the $5$-dimensional fields dual to glueballs/oddballs are excited from the background. The action describing scalar, vector, and tensor glueballs/oddballs in the string frame is
\begin{alignat}{3}
   & S_{g}^{s}=-c_g \int \mathrm{d}^{5} x \sqrt{-g_{s}} \mathrm{e}^{-p\Phi}
  \Bigg\{
   &                                                                        & \bigg[ \frac{1}{2}{g^s}^{M N}\partial_{M} \mathscr{S}(z) \partial_{N} \mathscr{S}+\frac{1}{2} \mathrm{e}^{-c_{\text{r.m.}}\Phi} M_{\mathscr{S}, 5}^{2} \mathscr{S}^{2} \bigg]
  \nonumber                                                                                                                                                                                                                                                                                                                                                                                                                                                                                                                         \\
   &                                                                        &                                                                                                                                                                               & +\bigg[ \frac{1}{4} {g^{s}}^{M \widetilde{M} }{g^{s}}^{N \widetilde{N} }\left(\partial_{M} \mathscr{V}_{N}-\partial_{N} \mathscr{V}_{M}\right) \left(\partial_{\widetilde{M}} \mathscr{V}_{\widetilde{N}}-\partial_{\widetilde{N}} \mathscr{V}_{\widetilde{M}}\right)
  \nonumber                                                                                                                                                                                                                                                                                                                                                                                                                                                                                                                         \\
   &                                                                        &                                                                                                                                                                               & \qquad +\frac{1}{2} \mathrm{e}^{-c_{\text{r.m.}}\Phi} M_{\mathscr{V}, 5}^{2} \mathscr{V}^{2}\bigg]
  \nonumber                                                                                                                                                                                                                                                                                                                                                                                                                                                                                                                         \\
   &                                                                        &                                                                                                                                                                               & +\bigg[
  \frac{1}{2} \nabla_{L} \mathscr{T}_{M N} \nabla^{L} \mathscr{T}^{M N}- \nabla_{L} \mathscr{T}^{L M} \nabla^{N} \mathscr{T}_{N M}
  \nonumber                                                                                                                                                                                                                                                                                                                                                                                                                                                                                                                         \\
   &                                                                        &                                                                                                                                                                               & \qquad + \nabla_{M} \mathscr{T}^{M N} \nabla_{N} \mathscr{T}-\frac{1}{2}\nabla_{M} \mathscr{T} \nabla^{M} \mathscr{T}
  \nonumber                                                                                                                                                                                                                                                                                                                                                                                                                                                                                                                         \\
   &                                                                        &                                                                                                                                                                               & \qquad +\frac{1}{2} \mathrm{e}^{-c_{\text{r.m.}}\Phi} M_{\mathscr{T}, 5}^{2}\left(\mathscr{T}^{M N} h_{M N}-\mathscr{T}^{2}\right)
  \bigg]
  \nonumber                                                                                                                                                                                                                                                                                                                                                                                                                                                                                                                         \\
   &                                                                        &                                                                                                                                                                               & +\text{terms for high spin fields (spin $S \geqslant 3$)}
  \Bigg\},
  \label{action_glueb}
\end{alignat}
where $s$ denotes the string frame, $c_g$ describes the coupling strength of glueballs part in the whole theory. The fields $\mathscr{S}$, $\mathscr{V}_{M}$, and $\mathscr{T}_{M N}$ are $5$-dimensional fields that are dual to scalar glueball, vector glueball, and spin-2 glueball operators respectively. $\mathscr{T}={g^{s}}^{M N} {\mathscr{T}}_{M N}$, and ${\mathscr{T}}_{M N}$ satisfies the following constraints
\begin{align}
  \nabla_{M} {\mathscr{T}}^{M N}=0, \quad {\mathscr{T}}=0, \quad {\mathscr{T}}_{\mu \nu}=\frac{1}{z^2}\mathrm{e}^{2 A_{s}} \mathscr{T}_{\mu \nu}, \quad {\mathscr{T}}_{M z}=0.
  \label{tens_glueb_const}
\end{align}
As in Ref. \cite{Chen:2015zhh}, the parameter $p$ is introduced to make a distinction between glueballs (oddballs) with different P\hyp{}parity:
\begin{align}
  \left\{
  \begin{aligned}
     & p=1,\quad \,\,\,\,\, \text{for even parity,}
    \\
     & p=-1,\quad \text{for odd parity.}
  \end{aligned}
  \right.
  \label{val_of_p}
\end{align}
Also we introduce a $z$ dependent modified $5$-dimensional mass:
\begin{align}
   & M_5^2(z)=\mathrm{e}^{-c_{\text{r.m.}}\Phi} M_5^2,
  \label{running_5D_mass}
\end{align}
where $c_{\text{r.m.}}$ is a constant. The $M_5^2$ is listed in Table \ref{twotrigluon-glueball} given by the $\text{AdS}_{5}/\text{CFT}_{4}$ correspondence dictionary. The $\text{AdS}_{5}/\text{CFT}_{4}$ duality gives one-to-one correspondence between $4$-dimensional operators in the ${\cal N}=4$ Super Yang-Mills theory and the spectrum of the type \RNum{2}B string theory on $AdS_{5}\times S^{5}$. Based on the AdS/CFT dictionary, the conformal dimension of a $p$-form operator at the ultraviolet (UV) boundary is related to the $5$-dimensional mass square $M_5^{2}$ of its dual field in the bulk as follows \cite{Maldacena:1997re,Gubser:1998bc,Witten:1998qj} :
\begin{equation}
  M_5^2=(\Delta-p)(\Delta+p-4).  \label{Eq-m5}
\end{equation}

\subsection{Glueballs and oddballs}
\label{glueb_and_oddb}
In the bottom-up holographic QCD models, one can expect a more general correspondence, i.e. each $4$-dimensional operator ${\cal O}(x)$ corresponds to a
$5$-dimensional field $O(x,z)$ in the bulk theory. To investigate the glueball spectra, we consider the lowest dimension operators with the
corresponding quantum numbers defined in the field theory living on the $4$-dimensional boundary. We show the C-even/odd glueball and oddball operators and their
corresponding $5$-dimensional mass square in Table \ref{twotrigluon-glueball}.

\begin{table}[htb]
  \begin{center}
    \begin{tabular}{|c|c|c|c|c|c|}
      \hline
      $J^{PC}$ & $4$-dimensional operator: $\mathscr{O}(x)$                                                                                                                                                                                                                           & $\Delta$ & $p$ & $M_5^2$ \\
      \hline
      $0^{++}$ & $Tr(G^2)=\vec{E}^{a} \cdot \vec{E}^{a}-\vec{B}^{a} \cdot \vec{B}^{a}$                                                                                                                                                                                                & 4        & 0   & 0       \\
      \hline
      $0^{-+}$ & $Tr(G\tilde{G})=\vec{E}^{a} \cdot \vec{B}^{a}$                                                                                                                                                                                                                       & 4        & 0   & 0       \\
      \hline
      $0^{+-}$ & $\operatorname{Tr}\left(\left\{\left(D_{\tau} G_{\mu \nu}\right),\left(D_{\tau} G_{\rho \nu}\right)\right\}\left(D_{\mu} G_{\rho \alpha}\right)\right)$                                                                                                              & 9        & 0   & 45      \\
      \hline
      $0^{--}$ & $\operatorname{Tr}\left(\left\{\left(D_{\tau} G_{\mu \nu}\right),\left(D_{\tau} G_{\rho \nu}\right)\right\}\left(D_{\mu} \tilde{G}_{\rho \alpha}\right)\right)$                                                                                                      & 9        & 0   & 45      \\
      \hline
      $1^{-+}$ & $f^{a b c} \partial_{\mu}\left[G_{\mu \nu}^{a}\right]\left[G_{v \rho}^{b}\right]\left[G_{\rho \alpha}^{c}\right]$, $f^{a b c} \partial_{\mu}\left[G_{\mu \nu}^{a}\right]\left[\tilde{G}_{v \rho}^{b}\right]\left[\tilde{G}_{\rho \alpha}^{c}\right]$,                & 7        & 1   & 24      \\
               & $f^{a b c} \partial_{\mu}\left[\tilde{G}_{\mu \nu}^{a}\right]\left[G_{v \rho}^{b}\right]\left[\tilde{G}_{\rho \alpha}^{c}\right]$, $f^{a b c} \partial_{\mu}\left[\tilde{G}_{\mu \nu}^{a}\right]\left[\tilde{G}_{v \rho}^{b}\right]\left[G_{\rho \alpha}^{c}\right]$ &          &     &         \\
      \hline
      $1^{+-}$ & $d^{a b c}\left(\vec{E}_{a} \cdot \vec{E}_{b}\right) \vec{B}_{c}$                                                                                                                                                                                                    & 6        & 1   & 15      \\
      \hline
      $1^{--}$ & $d^{a b c}\left(\vec{E}_{a} \cdot \vec{E}_{b}\right) \vec{E}_{c}$                                                                                                                                                                                                    & 6        & 1   & 15      \\
      \hline
      $2^{++}$ & $E_{i}^{a}E_{j}^{a}-B_{i}^{a}B_{j}^{a}-trace$                                                                                                                                                                                                                        & 4        & 2   & 4       \\
      \hline
      $2^{-+}$ & $E_{i}^{a}B_{j}^{a}+B_{i}^{a}E_{j}^{a}-trace$                                                                                                                                                                                                                        & 4        & 2   & 4       \\
      \hline
      $2^{+-}$ & $d^{a b c} \mathcal{S}\left[E_{a}^{i}\left(\vec{E}_{b} \times \vec{B}_{c}\right)^{j}\right]$                                                                                                                                                                         & 6        & 2   & 16      \\
      \hline
      $2^{--}$ & $d^{a b c} \mathcal{S}\left[B_{a}^{i}\left(\vec{E}_{b} \times \vec{B}_{c}\right)^{j}\right]$                                                                                                                                                                         & 6        & 2   & 16      \\
      \hline
      $3^{+-}$ & $d^{a b c} \mathcal{S}\left[B_{a}^{i} B_{b}^{j} B_{c}^{k}\right]$                                                                                                                                                                                                    & 6        & 3   & 15      \\
      \hline
      $3^{--}$ & $d^{a b c} \mathcal{S}\left[E_{a}^{i} E_{b}^{j} E_{c}^{k}\right]$                                                                                                                                                                                                    & 6        & 3   & 15      \\
      \hline
    \end{tabular}
    \caption{$5$-dimensional mass square of C-even glueballs and C-odd oddballs. The operators are taken from Refs. \cite{Brower:2000rp,Pimikov:2017bkk,Tang:2015twt,Chen:2021cjr}.}
    \label{twotrigluon-glueball}
  \end{center}
\end{table}

The lowest dimension gauge invariant three-gluon currents that couple to the exotic $0^{+-}$ and $0^{--}$ glueballs are constructed in Ref. \cite{Pimikov:2017bkk}:
\begin{align}
   & j_{\alpha}^{0^{+-}}(x)=g_{s}^{3} \operatorname{Tr}\left(\left\{\left(D_{\tau} G_{\mu \nu}(x)\right),\left(D_{\tau} G_{\rho \nu}(x)\right)\right\}\left(D_{\mu} G_{\rho \alpha}(x)\right)\right),
  \label{current_0_p_m}                                                                                                                                                                                       \\
   & j_{\alpha}^{0^{--}}(x)=g_{s}^{3} \operatorname{Tr}\left(\left\{\left(D_{\tau} G_{\mu \nu}(x)\right),\left(D_{\tau} G_{\rho \nu}(x)\right)\right\}\left(D_{\mu} \tilde{G}_{\rho \alpha}(x)\right)\right).
  \label{current_0_m_m}
\end{align}

For trigluon glueball $1^{-+}$, and $2^{+-}$, the currents that match the unconventional quantum number and satisfy the constraints of the gauge invariance are given in Refs. \cite{Tang:2015twt}:
\begin{align}
   & j_{\alpha}^{1^{-+}, A}(x) =g_{s}^{3} f^{a b c} \partial_{\mu}\left[G_{\mu \nu}^{a}(x)\right]\left[G_{v \rho}^{b}(x)\right]\left[G_{\rho \alpha}^{c}(x)\right],
  \nonumber                                                                                                                                                                            \\
   & j_{\alpha}^{1^{-+}, B}(x) =g_{s}^{3} f^{a b c} \partial_{\mu}\left[G_{\mu \nu}^{a}(x)\right]\left[\tilde{G}_{v \rho}^{b}(x)\right]\left[\tilde{G}_{\rho \alpha}^{c}(x)\right],
  \nonumber                                                                                                                                                                            \\
   & j_{\alpha}^{1^{-+}}, C_{(x)} =g_{s}^{3} f^{a b c} \partial_{\mu}\left[\tilde{G}_{\mu \nu}^{a}(x)\right]\left[G_{v \rho}^{b}(x)\right]\left[\tilde{G}_{\rho \alpha}^{c}(x)\right],
  \nonumber                                                                                                                                                                            \\
   & j_{\alpha}^{1^{-+}, D}(x) =g_{s}^{3} f^{a b c} \partial_{\mu}\left[\tilde{G}_{\mu \nu}^{a}(x)\right]\left[\tilde{G}_{v \rho}^{b}(x)\right]\left[G_{\rho \alpha}^{c}(x)\right],
  \label{current_1_m_p}
\end{align}
and
\begin{align}
   & j_{\mu \alpha}^{2^{+-}, A}(x)=g_{s}^{3} d^{a b c}\left[G_{\mu \nu}^{a}(x)\right]\left[G_{v \rho}^{b}(x)\right]\left[G_{\rho \alpha}^{c}(x)\right],
  \nonumber                                                                                                                                                             \\
   & j_{\mu \alpha}^{2^{+-}, B}(x)=g_{s}^{3} d^{a b c}\left[G_{\mu \nu}^{a}(x)\right]\left[\tilde{G}_{v \rho}^{b}(x)\right]\left[\tilde{G}_{\rho \alpha}^{c}(x)\right],
  \nonumber                                                                                                                                                             \\
   & j_{\mu \alpha}^{2^{+-}, C}(x)=g_{s}^{3} d^{a b c}\left[\tilde{G}_{\mu \nu}^{a}(x)\right]\left[G_{v \rho}^{b}(x)\right]\left[\tilde{G}_{\rho \alpha}^{c}(x)\right],
  \nonumber                                                                                                                                                             \\
   & j_{\mu \alpha}^{2^{+-}, D}(x)=g_{s}^{3} d^{a b c}\left[\tilde{G}_{\mu \nu}^{a}(x)\right]\left[\tilde{G}_{v \rho}^{b}(x)\right]\left[G_{\rho \alpha}^{c}(x)\right],
  \label{current_2_p_m}
\end{align}
where $d^{abc}$ stands for the totally symmetric $SU_c(3)$ structure constant and $g^t_{\alpha \beta}= g_{\alpha \beta}- \partial_\alpha \partial_\beta/\partial^2$.

\subsection{Equation of motion for scalar, vector and tensor glueballs/oddballs}

From the $5$-dimensional action for the glueball/oddball in the string frame Eq. (\ref{action_glueb}), we can derive the equation of motion for the glueballs.
The equation of motion for the scalar glueballs $\mathscr{S}$  is given as:
\begin{align}
   & -z^3 {\mathrm{e}}^{-(3A_s-p\Phi)}\partial_z\left[\frac{1}{z^3}{\mathrm{e}}^{3A_s-p\Phi}\partial_z{\mathscr{S}_n}\right]
  \nonumber                                                                                                                                         \\
   & +\frac{1}{z^2}{\mathrm{e}}^{2A_s} \mathrm{e}^{-c_{\text{r.m.}}\Phi} M_{\mathscr{S}, 5}^{2}{\mathscr{S}_n}=m_{\mathscr{S},n}^2 {\mathscr{S}_n}.
\end{align}
Via the substitution
\begin{align}
   & \mathscr{S}_n \rightarrow z^{\frac{3}{2}}\mathrm{e}^{-\frac{1}{2}(3A_s-p\Phi)}\mathscr{S}_n,
  \label{field_substitution_scalar}
\end{align}
the equation can be brought into Schr\"{o}dinger\hyp{}like equation
\begin{eqnarray}
  -{\mathscr{S}_n}^{''}+V_{\mathscr{S}} {\mathscr{S}_n}=m_{\mathscr{G},n}^2 {\mathscr{S}_n},
  \label{EOM-glueball}
\end{eqnarray}
with the $5$-dimensional effective Schr\"{o}dinger potential
\begin{equation}
  V_{\mathscr{S}}=\frac{3A_s^{''}+\frac{3}{z^2}-p\Phi^{''}}{2}+\frac{\left[3A_s^{'}-\frac{3}{z}-p\Phi^{'}\right]^2}{4}+\frac{1}{z^2}e^{2A_s} \mathrm{e}^{-c_{\text{r.m.}}\Phi} M_{\mathscr{S}, 5}^{2}.
  \label{potential-glueball-s}
\end{equation}

The equation of motion for the vector glueballs $\mathscr{V}_{M}$ is given as:
\begin{eqnarray}
  -z\mathrm{e}^{-(A_s-p\Phi)}\partial_z\left[\frac{1}{z}\mathrm{e}^{A_s-p\Phi}\partial_z\mathscr{V}_n\right]+\frac{1}{z^2}\mathrm{e}^{2A_s} \mathrm{e}^{-c_{\text{r.m.}}\Phi} M_{\mathscr{V}, 5}^{2}\mathscr{V}_n=m_{\mathscr{V},n}^2 \mathscr{V}_n.
\end{eqnarray}
Via the substitution
\begin{align}
   & \mathscr{V}_n \rightarrow z^{\frac{1}{2}}e^{-\frac{1}{2}(A_s-p\Phi)}\mathscr{V}_n,
  \label{field_substitution_vector}
\end{align}
the equation can be brought into Schr\"{o}dinger\hyp{}like equation
\begin{eqnarray}
  -\mathscr{V}_n^{''}+V_{\mathscr{V}} \mathscr{V}_n=m_{\mathscr{V},n}^2 \mathscr{V}_n,
\end{eqnarray}
with the $5$-dimensional effective Schr\"{o}dinger potential
\begin{equation}
  V_{\mathscr{V}}=\frac{A_s^{''}+\frac{1}{z^2}-p\Phi^{''}}{2}+\frac{\left[A_s^{'}-\frac{1}{z}-p\Phi^{'}\right]^2}{4}+\frac{1}{z^2}\mathrm{e}^{2A_s} \mathrm{e}^{-c_{\text{r.m.}}\Phi} M_{\mathscr{V},5}^2.
  \label{potential-glueball-v}
\end{equation}

The equation of motion for the spin-2 glueballs $\mathscr{T}_{M N}$ is given as
\begin{eqnarray}
  -z^3 \mathrm{e}^{-(3A_s-p\Phi)}\partial_z\left[\frac{1}{z^3}\mathrm{e}^{3A_s-p\Phi}\partial_z\mathscr{T}_n\right]+\frac{1}{z^2}\mathrm{e}^{2A_s} \mathrm{e}^{-c_{\text{r.m.}}\Phi} M_{\mathscr{T}, 5}^{2}\mathscr{T}_n=m_{\mathscr{T},n}^2\mathscr{T}_n.
\end{eqnarray}
Via the substitution
\begin{align}
   & \mathscr{T}_n \rightarrow z^{\frac{3}{2}}\mathrm{e}^{-\frac{1}{2}(3A_s-p\Phi)}\mathscr{T}_n,
  \label{field_substitution_s_2}
\end{align}
the equation can be brought into Schr\"{o}dinger\hyp{}like equation
\begin{eqnarray}
  -\mathscr{T}_n^{''}+V_{\mathscr{T}} \mathscr{T}_n=m_{\mathscr{T},n}^2 \mathscr{T}_n,
\end{eqnarray}
with the $5$-dimensional effective Schr\"{o}dinger potential
\begin{equation}
  V_{\mathscr{T}}=\frac{3A_s^{''}+\frac{3}{z^2}-p\Phi^{''}}{2}+\frac{\left[3A_s^{'}-\frac{3}{z}-p\Phi^{'}\right]^2}{4}+\frac{1}{z^2}\mathrm{e}^{2A_s} \mathrm{e}^{-c_{\text{r.m.}}\Phi} M_{\mathscr{T}, 5}^{2}.
  \label{potential-glueball-s2_t}
\end{equation}

According to Ref. \cite{Karch:2006pv}, the equation of motion for the high spin glueballs $\mathscr{H}_{M_{1}M_{2}\cdots M_{S}}$, the spin $S$ of which are larger than $2$, is given as
\begin{align}
   & -z^{2S-1} \mathrm{e}^{-\left[\left(2S-1\right)A_s-p\Phi\right]}\partial_z\left[\frac{1}{z^{2S-1}}\mathrm{e}^{\left(2S-1\right)A_s-p\Phi}\partial_z\mathscr{H}_n\right]
  \nonumber                                                                                                                                                                 \\
   & +\frac{1}{z^2}\mathrm{e}^{2A_s} \mathrm{e}^{-c_{\text{r.m.}}\Phi} M_{\mathscr{H}, 5}^{2}\mathscr{H}_n=m_{\mathscr{H},n}^2\mathscr{H}_n,
  \label{eom_of_high_spin_field}
\end{align}
where $S \geqslant 3$.
Via the substitution
\begin{align}
   & \mathscr{H}_n \rightarrow z^{\frac{2S-1}{2}}\mathrm{e}^{-\frac{1}{2}\left[\left(2S-1\right)A_s-p\Phi\right]}\mathscr{H}_n,
  \label{field_substitution_high_spin}
\end{align}
the equation can be brought into Schr\"{o}dinger\hyp{}like equation
\begin{eqnarray}
  -\mathscr{H}_n^{''}+V_{\mathscr{H}} \mathscr{H}_n=m_{\mathscr{H},n}^2 \mathscr{H}_n,
\end{eqnarray}
with the $5$-dimensional effective Schr\"{o}dinger potential
\begin{equation}
  V_{\mathscr{H}}=\frac{\left(2S-1\right)A_s^{''}+\frac{2S-1}{z^2}-p\Phi^{''}}{2}+\frac{\left[\left(2S-1\right)A_s^{'}-\frac{2S-1}{z}-p\Phi^{'}\right]^2}{4}+\frac{1}{z^2}\mathrm{e}^{2A_s} \mathrm{e}^{-c_{\text{r.m.}}\Phi} M_{\mathscr{H}, 5}^{2}.
  \label{potential_glueball_high_spin}
\end{equation}

\subsection{Numerical results of glueballs/oddballs spectra}

We calculate the glueballs spectra using five different holographic models defined in last section. We list the parameters used for calculating the glueballs spectra below.

\subsubsection{Model \texorpdfstring{\RNum{1}}{1} and \texorpdfstring{\RNum{2}}{2}}
\label{glueballs_spectra_model_1_and_2}

In model \RNum{1} and model \RNum{2}, we choose the parameter $a=0.6032 \mathrm{GeV}^2$. Firstly, we don't consider the distinction between glueballs (oddballs) with different P\hyp{}parity and don't introduce $z$ dependent modified $5$-dimensional masses, that means $p=1$ for even and odd parity, and $c_{\text{r.m.}}=0$. Then we calculate the glueballs/oddballs mass spectra in model \RNum{1} and \RNum{2}, which is denoted by "Model \RNum{1},\RNum{2}(O)" in Tab. \ref{tab_glueb_spect_original_holog_and_latti}. We find the calculation results of the masses of glueballs/oddballs, of which the $5$D mass square $M_5^2$ in Tab. \ref{twotrigluon-glueball} are large, are much heavier than the lattice data, as we mentioned in subsubsection \ref{glueballs_spectra_model_1_and_2}. That's why we introduce a $z$-dependent modified $5$-dimensional mass of glueball/oddball fields in Eq. (\ref{running_5D_mass}). The value of the constant $c_{\text{r.m.}}$ in Eq. (\ref{running_5D_mass}) is chosen as $\frac{3}{5}$, which means
\begin{align}
   & M_{5}^{2}(z) = {\mathrm{e}}^{-\frac{3}{5}\Phi} M_{5}^{2}, ~~~\text{model \RNum{1}, and \RNum{2}}.
  \label{5dmassIandII}
\end{align}

The results of glueballs/oddballs spectra are denoted by "Model \RNum{1},\RNum{2}" in Tab. \ref{tab_glueb_spect_holog_and_latti}.

Note that the $5$-dimensional field $\Phi$ and $\phi$ are different, the relationship between them is Eq. (\ref{new_dilaton}):
\begin{align}
   & \phi=\sqrt{\frac{8}{3}}\Phi.
  \nonumber
\end{align}

\subsubsection{Model \texorpdfstring{\RNum{3}}{3} and \texorpdfstring{\RNum{4}}{4}}
\label{glueballs_spectra_model_3_and_4}

In model \RNum{3} and model \RNum{4}, we choose the parameter $b=1.760 \mathrm{GeV}^2$. The value of the constant $c_{\text{r.m.}}$ in Eq. (\ref{running_5D_mass}) is chosen as $\frac{1}{2}$, which means
\begin{align}
   & M_{5}(z)^{2} = {\mathrm{e}}^{-\frac{1}{2}\Phi} M_{5}^{2}, ~~~\text{model \RNum{3} and \RNum{4}}.
  \label{5dmassIIIandIV}
\end{align}

The results of glueballs/oddballs spectra is denoted as "Model \RNum{3},\RNum{4}(1)" in Tab. \ref{tab_glueb_spect_holog_and_latti}.

In Ref. \cite{Chen:2015zhh}, the authors also use model \RNum{3} to calculate the glueballs spectra. There they use the parameter $b=\frac{2 \sqrt{6}}{3} \mathrm{GeV}^2$ \footnote{
  Please remember Eq. (\ref{new_dilaton}). This value of $b$ means
  \begin{equation}
    \Phi(z)=\left(1 \mathrm{GeV}^2\right) z^2.
    \label{input_1511_07018}
  \end{equation}
} and $c_{\text{r.m.}}=\frac{2}{3}$. We also calculate the glueballs spectra using these parameters and list the results denoted by "Model \RNum{3},\RNum{4}(2)" in the Tab. \ref{tab_glueb_spect_holog_and_latti}.

\subsubsection{Model \texorpdfstring{\RNum{5}}{5}}
\label{glueballs_spectra_model_5}

In model \RNum{5}, we choose the parameter $d=0.2 \mathrm{GeV}^2$. The value of the constant $c_{\text{r.m.}}$ in Eq. (\ref{running_5D_mass}) is chosen as $\frac{1}{3}$, which means
\begin{equation}
  M_{5}(z)^{2} =  {\mathrm{e}}^{-\frac{1}{3}\Phi} M_{5}^{2}, ~~~\text{model \RNum{5}}.
  \label{5dmassV}
\end{equation}

The results are denoted by "Model \RNum{5}" in Tab. \ref{tab_glueb_spect_holog_and_latti}.

The corresponding results for glueballs and oddballs spectra are also shown in Fig. \ref{fig_glueb_spect_holog_and_latti_C_even} and Fig. \ref{fig_glueb_spect_holog_and_latti_C_odd}, respectively.

\subsubsection{Compare results with lattice QCD, QCD sum rule and \texorpdfstring{$pp$}{pp} high energy scattering}
\label{compare_with_lattice_data}

\begin{figure}
  \begin{center}
    \includegraphics[width=135mm,clip=true,keepaspectratio=true]{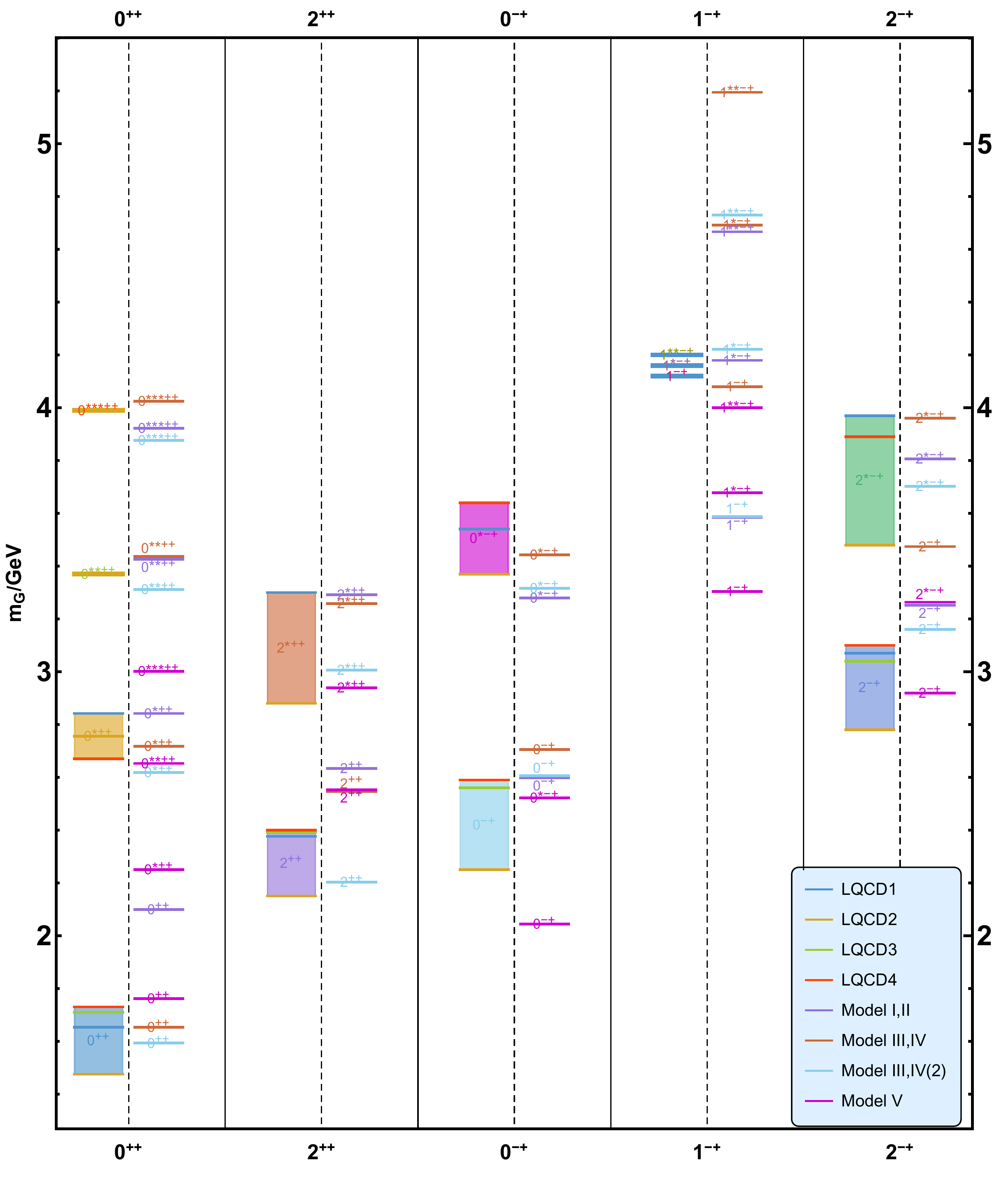}
  \end{center}
  \caption{The mass spectra of $J^{PC}$ ($C=1$) glueballs in the dynamical soft-wall model, compared with lattice data. This figure are split into five panels, that are divided by black solid lines. From left to right, the mass data in these panels belong to $0^{++}$ states, $2^{++}$ states, $0^{-+}$ states, $1^{-+}$ states, and $2^{-+}$ states respectively. In every panel, the black dashed line split it into two parts. The left one contains lattice data taken from Refs. \cite{Athenodorou:2020ani,Meyer:2004gx,Chen:2005mg,Morningstar:1999rf}. The steel blue lines, goldenrod lines, olive drab lines, orange red lines are lattice data taken from Ref. \cite{Athenodorou:2020ani}, Ref \cite{Meyer:2004gx}, Ref \cite{Chen:2005mg}, and Ref \cite{Morningstar:1999rf} respectively. The minimal value and maximal value of a set of discrete data that belongs to the same glueball state decide the positions of lower and upper bound of the bar in the figure respectively. The data in the right part are calculated in our holographic models. The medium purple lines, sienna lines, sky blue lines, and magenta lines are results from "Model \RNum{1},\RNum{2}", "Model \RNum{3},\RNum{4}(1)", "Model \RNum{3},\RNum{4}(2)", and "Model \RNum{5}" respectively.}
  \label{fig_glueb_spect_holog_and_latti_C_even}
\end{figure}

\begin{figure}
  \begin{center}
    \includegraphics[width=135mm,clip=true,keepaspectratio=true]{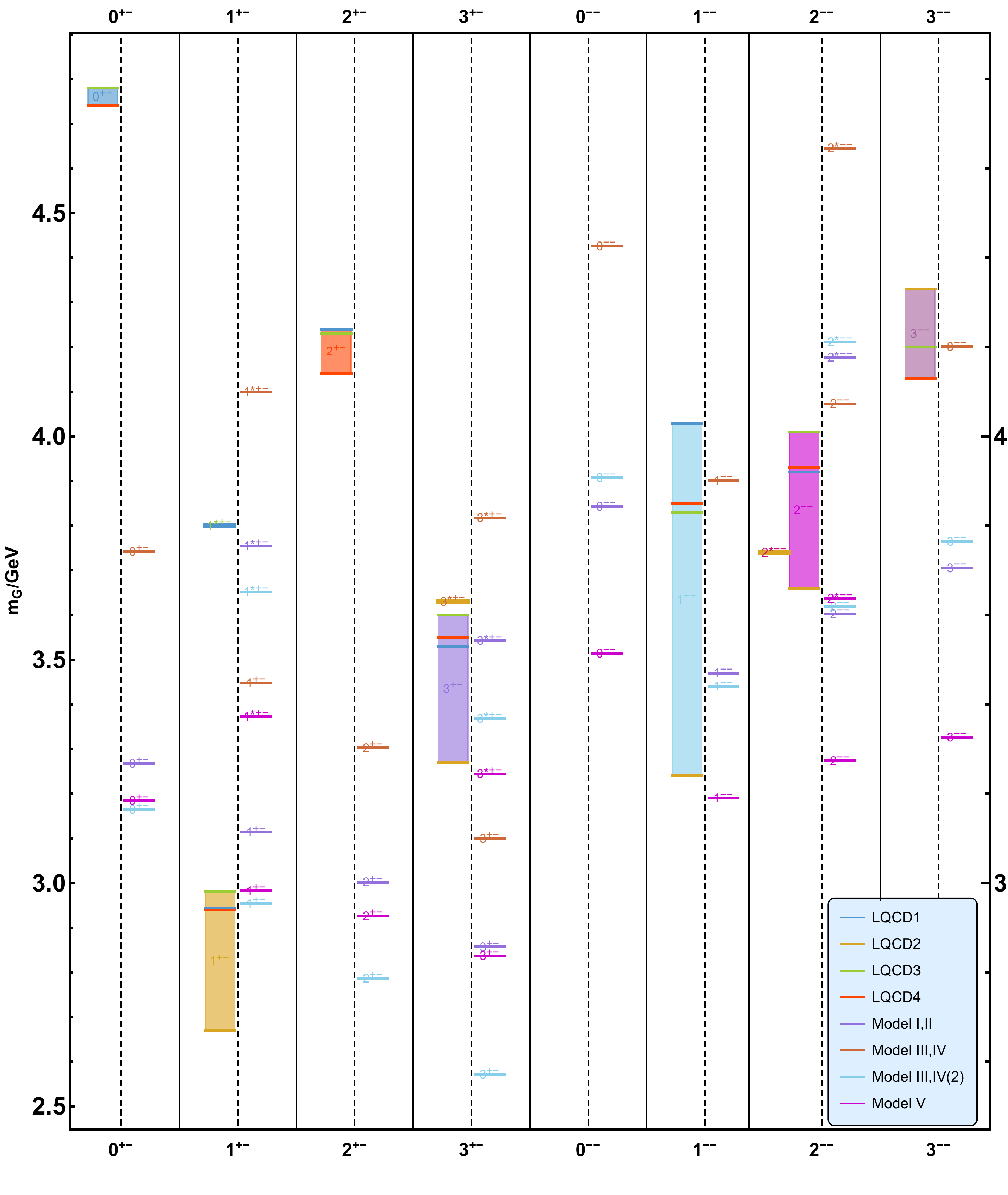}
  \end{center}
  \caption{The mass spectra of $J^{PC}$ ($C=-1$) oddballs in the dynamical soft-wall model, compared with lattice data. This figure are split into eight panels, that are divided by black solid lines. From left to right, the mass data in these panels belong to $0^{+-}$ states, $1^{+-}$ states, $2^{+-}$ states, $3^{+-}$ states, $0^{--}$ states, $1^{--}$ states, $2^{--}$ states, and $3^{--}$ states respectively. In every panel, the black dashed line split it into two parts. The left one contains lattice data taken from Refs. \cite{Athenodorou:2020ani,Meyer:2004gx,Chen:2005mg,Morningstar:1999rf}. The steel blue lines, goldenrod lines, olive drab lines, orange red lines are lattice data taken from Ref. \cite{Athenodorou:2020ani}, Ref \cite{Meyer:2004gx}, Ref \cite{Chen:2005mg}, and Ref \cite{Morningstar:1999rf} respectively. The minimal value and maximal value of a set of discrete data that belongs to the same oddball state decide the positions of lower and upper bound of the bar in the figure respectively. The data in the right part are calculated in our holographic models. The medium purple lines, sienna lines, sky blue lines, and magenta lines are results from "Model \RNum{1},\RNum{2}", "Model \RNum{3},\RNum{4}(1)", "Model \RNum{3},\RNum{4}(2)", and "Model \RNum{5}" respectively.}
  \label{fig_glueb_spect_holog_and_latti_C_odd}
\end{figure}

We summarize our holographic results of glueballs/oddballs spectra and then compare them with the results from lattice simulation and QCD sum rule in Tab. \ref{tab_glueb_spect_holog_and_latti}. To explicitly see the difference between results from holographic QCD models and those from lattice simulation, we also list results in Fig. \ref{fig_glueb_spect_holog_and_latti_C_even} for C-even glueballs, and in Fig. \ref{fig_glueb_spect_holog_and_latti_C_odd} for C-odd oddballs.

\begin{table}[htb!]
  \begin{tiny}
    \begin{center}
      \begin{tabular}{|c|c|c|c|c|c|c|}
        \hline
        $J^{PC}$    & LQCD1     & LQCD2           & LQCD3          & LQCD4            & QCDSR                   & Model \RNum{1},\RNum{2}(O) \\
        \hline
        $0^{++}$    & 1.653(26) & 1.475(30)(65)   & 1.710(50)(80)  & 1.730 (50) (80)  & $1.50 \pm 0.19$         & 2.099                      \\
        \hline
        $0^{*++}$   & 2.842(40) & 2.755(70)(120)  & --             & 2.670 (180)(130) & 2.0 - 2.1               & 2.842                      \\
        \hline
        $0^{**++}$  & --        & 3.370(100)(150) & --             & --               & --                      & 3.425                      \\
        \hline
        $0^{***++}$ & --        & 3.990(210)(180) & --             & --               & --                      & 3.922                      \\
        \hline
        $2^{++}$    & 2.376(32) & 2.150(30)(100)  & 2.390(30)(120) & 2.400 (25) (120) & $2.0 \pm 0.1$           & 8.831                      \\
        \hline
        $2^{*++}$   & 3.30(5)   & 2.880(100)(130) & --             & --               & 2.2 - 2.3               & 9.515                      \\
        \hline
        $0^{-+}$    & 2.561(40) & 2.250(60)(100)  & 2.560(35)(120) & 2.590 (40) (130) & $2.05 \pm 0.19$         & 2.099                      \\
        \hline
        $0^{*-+}$   & 3.54(8)   & 3.370(150)(150) & --             & 3.640 (60) (180) & 2.1 - 2.3               & 2.842                      \\
        \hline
        $1^{-+}$    & 4.12(8)   & --              & --             & --               & --                      & 20.675                     \\
        \hline
        $1^{*-+}$   & 4.16(8)   & --              & --             & --               & --                      & 21.405                     \\
        \hline
        $1^{**-+}$  & 4.20(9)   & --              & --             & --               & --                      & 22.093                     \\
        \hline
        $2^{-+}$    & 3.07(6)   & 2.780(50)(130)  & 3.040(40)(150) & 3.100 (30) (150) & --                      & 8.831                      \\
        \hline
        $2^{*-+}$   & 3.97(7)   & 3.480(140)(160) & --             & 3.890 (40) (190) & --                      & 9.515                      \\
        \hline
        $0^{+-}$    & --        & --              & 4.780(60)(230) & 4.740 (70) (230) & $9.2_{-1.4}^{+1.3}$     & 28.137                     \\
        \hline
        $1^{+-}$    & 2.944(42) & 2.670(65)(120)  & 2.980(30)(140) & 2.940 (30) (140) & $2.87_{-0.20}^{+0.17}$  & 16.457                     \\
        \hline
        $1^{*+-}$   & 3.80(6)   & --              & --             & --               & --                      & 17.176                     \\
        \hline
        $2^{+-}$    & 4.24(8)   & --              & 4.230(50)(200) & 4.140 (50) (200) & $2.85_{-0.20}^{+0.16}$  & 16.996                     \\
                    &           &                 &                &                  & $6.06\pm{0.13}$         &                            \\
        \hline
        $3^{+-}$    & 3.53(8)   & 3.270(90)(150)  & 3.600(40)(170) & 3.550 (40) (170) & $2.78_{-0.23}^{+0.18}$  & 16.491                     \\
        \hline
        $3^{*+-}$   & --        & 3.630(140)(160) & --             & --               & --                      & 17.212                     \\
        \hline
        $0^{--}$    & --        & --              & --             & --               & $6.8_{-1.2}^{+1.1}$     & 28.137                     \\
        \hline
        $1^{--}$    & 4.03(7)   & 3.240(330)(150) & 3.830(40)(190) & 3.850 (50) (190) & $3.29_{-0.32}^{+1.49}$  & 16.457                     \\
        \hline
        $2^{--}$    & 3.92(9)   & 3.660(130)(170) & 4.010(45)(200) & 3.930 (40) (190) & $3.16_{-0.23}^{+0.33}$  & 16.996                     \\
        \hline
        $2^{*--}$   & --        & 3.740(200)(170) & --             & --               & --                      & 17.717                     \\
        \hline
        $3^{--}$    & --        & 4.330(260)(200) & 4.200(45)(200) & 4.130 (90) (200) & $3.47_{-0.50}^{+\;\;?}$ & 16.491                     \\
        \hline
      \end{tabular}
    \end{center}
    \caption{The glueballs and oddballs mass spectra in the dynamical soft-wall model \RNum{1} and \RNum{2} without making a distinction between glueballs (oddballs) with different $P$\hyp{}parity and introducing $z$ dependent modified $5$-dimensional masses, compared with results from lattice QCD and QCD sum rule. The units of all the data in the table are $\mathrm{GeV}$. The lattice data in the column "LQCD1", column "LQCD2", column "LQCD3", and column "LQCD4" are taken from Ref. \cite{Athenodorou:2020ani}, Ref \cite{Meyer:2004gx}, Ref \cite{Chen:2005mg}, and Ref \cite{Morningstar:1999rf} respectively. The QCD sum rule results are take from Refs. \cite{Narison:1996fm,Pimikov:2017bkk,Chen:2021cjr,Tang:2015twt}. Here we also list the data predicted by the single pole (SP) and dipole (DP) Regge model \cite{Szanyi:2019kkn}: using the SP Regge model, the predicted mass for $2^{++}$ glueball is $1.747 \mathrm{GeV}$; using the DP Regge model, the predicted masses for $2^{++}$ glueball and $3^{--}$ oddall are $1.758 \mathrm{GeV}$ and $3.001 \mathrm{GeV}$ respectively.}
    \label{tab_glueb_spect_original_holog_and_latti}
  \end{tiny}
\end{table}

\begin{table}[htb!]
  \begin{tiny}
    \begin{center}
      \begin{tabular}{|c|c|c|c|c|c|c|}
        \hline
        $J^{PC}$    & LQCD1-4                           & QCDSR                   & Model \RNum{1},\RNum{2} & Model \RNum{3},\RNum{4}(1) & Model \RNum{3},\RNum{4}(2) & Model \RNum{5} \\
        \hline
        $0^{++}$    & 1.475(30)(65) - 1.730(50)(80)     & $1.50 \pm 0.19$         & 2.099                   & 1.653                      & 1.593                      & 1.761          \\
        \hline
        $0^{*++}$   & 2.670 (180)(130) - 2.842(40)      & 2.0 - 2.1               & 2.842                   & 2.718                      & 2.618                      & 2.251          \\
        \hline
        $0^{**++}$  & 3.370(100)(150)                   & --                      & 3.425                   & 3.437                      & 3.311                      & 2.653          \\
        \hline
        $0^{***++}$ & 3.990(210)(180)                   & --                      & 3.922                   & 4.024                      & 3.877                      & 3.001          \\
        \hline
        $2^{++}$    & 2.150(30)(100) - 2.400(25)(120)   & $2.0 \pm 0.1$           & 2.633                   & 2.546                      & 2.203                      & 2.553          \\
        \hline
        $2^{*++}$   & 2.880(100)(130) - 3.30(5)         & 2.2 - 2.3               & 3.292                   & 3.258                      & 3.006                      & 2.939          \\
        \hline
        $0^{-+}$    & 2.250(60)(100) - 2.590(40)(130)   & $2.05 \pm 0.19$         & 2.599                   & 2.705                      & 2.606                      & 2.043          \\
        \hline
        $0^{*-+}$   & 3.370(150)(150) - 3.640(60)(180)  & 2.1 - 2.3               & 3.280                   & 3.443                      & 3.317                      & 2.522          \\
        \hline
        $1^{-+}$    & 4.12(8)                           & --                      & 3.585                   & 4.079                      & 3.588                      & 3.304          \\
        \hline
        $1^{*-+}$   & 4.16(8)                           & --                      & 4.180                   & 4.692                      & 4.221                      & 3.678          \\
        \hline
        $1^{**-+}$  & 4.20(9)                           & --                      & 4.667                   & 5.195                      & 4.730                      & 4.000          \\
        \hline
        $2^{-+}$    & 2.780(50)(130) - 3.100(30)(150)   & --                      & 3.252                   & 3.474                      & 3.161                      & 2.919          \\
        \hline
        $2^{*-+}$   & 3.480(140)(160) - 3.97(7)         & --                      & 3.806                   & 3.960                      & 3.703                      & 3.263          \\
        \hline
        $0^{+-}$    & 4.740 (70) (230) - 4.780(60)(230) & $9.2_{-1.4}^{+1.3}$     & 3.268                   & 3.742                      & 3.165                      & 3.184          \\
        \hline
        $1^{+-}$    & 2.670(65)(120) - 2.980(30)(140)   & $2.87_{-0.20}^{+0.17}$  & 3.114                   & 3.448                      & 2.954                      & 2.982          \\
        \hline
        $1^{*+-}$   & 3.80(6)                           & --                      & 3.755                   & 4.099                      & 3.652                      & 3.373          \\
        \hline
        $2^{+-}$    & 4.140 (50) (200) - 4.24(8)        & $2.85_{-0.20}^{+0.16}$, & 3.002                   & 3.303                      & 2.786                      & 2.926          \\
                    &                                   & $6.06\pm{0.13}$         &                         &                            &                            &                \\
        \hline
        $3^{+-}$    & 3.270(90)(150) - 3.600(40)(170)   & $2.78_{-0.23}^{+0.18}$  & 2.857                   & 3.100                      & 2.572                      & 2.837          \\
        \hline
        $3^{*+-}$   & 3.630(140)(160)                   & --                      & 3.542                   & 3.818                      & 3.369                      & 3.244          \\
        \hline
        $0^{--}$    & --                                & $6.8_{-1.2}^{+1.1}$     & 3.843                   & 4.426                      & 3.907                      & 3.514          \\
        \hline
        $1^{--}$    & 3.240(330)(150) - 4.03(7)         & $3.29_{-0.32}^{+1.49}$  & 3.470                   & 3.901                      & 3.441                      & 3.190          \\
        \hline
        $2^{--}$    & 3.660(130)(170) - 4.010(45)(200)  & $3.16_{-0.23}^{+0.33}$  & 3.602                   & 4.073                      & 3.619                      & 3.273          \\
        \hline
        $2^{*--}$   & 3.740(200)(170)                   & --                      & 4.177                   & 4.645                      & 4.211                      & 3.637          \\
        \hline
        $3^{--}$    & 4.130(90)(200) - 4.330(260)(200)  & $3.47_{-0.50}^{+\;\;?}$ & 3.705                   & 4.201                      & 3.765                      & 3.326          \\
        \hline
      \end{tabular}
    \end{center}
    \caption{The glueballs and oddballs mass spectra in the dynamical soft-wall model, compared with results from lattice QCD and QCD sum rule. The units of all the data in the table are $\mathrm{GeV}$. The lattice data are taken from Refs. \cite{Athenodorou:2020ani,Meyer:2004gx,Chen:2005mg,Morningstar:1999rf}. The QCD sum rule results are take from Refs. \cite{Narison:1996fm,Pimikov:2017bkk,Chen:2021cjr,Tang:2015twt}. Here we also list the data predicted by the SP and DP Regge model \cite{Szanyi:2019kkn}: using the SP Regge model, the predicted mass for $2^{++}$ glueball is $1.747 \mathrm{GeV}$; using the DP Regge model, the predicted masses for $2^{++}$ glueball and $3^{--}$ oddall are $1.758 \mathrm{GeV}$ and $3.001 \mathrm{GeV}$ respectively.}
    \label{tab_glueb_spect_holog_and_latti}
  \end{tiny}
\end{table}

In the framework of holography, the states $J^{PC}$ with the same angular momentum $J$ and the same C\hyp{}parity share the same operator. Thus, they have the same dimension and $5$-dimensional mass, and the mass splitting for different P\hyp{}parity states is realized by $\mathrm{e}^{-p\Phi}$ in Eq. (\ref{action_glueb}). The states $J^{PC}$ with the same angular momentum $J$ and the same P\hyp{}parity but different C\hyp{}parity have different operators. Thus, they have different $5$-dimensional masses, which naturally induces the mass splitting for different C\hyp{}parity states. From the results in Table \ref{tab_glueb_spect_holog_and_latti}, Fig.\ref{fig_glueb_spect_holog_and_latti_C_even}, and Fig. \ref{fig_glueb_spect_holog_and_latti_C_odd}, we can see that with only 2 parameters, the model predictions on glueballs/oddballs spectra in general are in good agreement with lattice results except two oddballs $0^{+-}$ and $2^{+-}$. Here we also would like to mention that the data predicted by the single pole (SP) and dipole (DP) Regge model \cite{Szanyi:2019kkn} to fit the high energy $pp$ scattering: using the SP Regge model, the predicted mass for $2^{++}$ glueball is $1.747 \mathrm{GeV}$; using the DP Regge model, the predicted masses for $2^{++}$ glueball and $3^{--}$ oddall are $1.758 \mathrm{GeV}$ and $3.001 \mathrm{GeV}$ respectively. These predicted values are a little bit lower than the results predicted from holography but still in reasonable regions. It might indicate that the mass $1.747 \mathrm{GeV}$/$1.758 \mathrm{GeV}$ $2^{++}$ glueball and mass $3.001 \mathrm{GeV}$ $3^{--}$ oddball are hybrid glueball/oddball states mixing with quark states.

\section{Equation of state}
\label{eos}

With parameters used to calculate the glueballs/oddballs spectra listed in Table \ref{tab_glueb_spect_holog_and_latti}, we check the corresponding thermodynamical properties of the system.

\subsection{Model \texorpdfstring{\RNum{1}}{1} and \texorpdfstring{\RNum{2}}{2}}

\begin{figure}[htb]
  \includegraphics[width=70mm,clip=true,keepaspectratio=true]{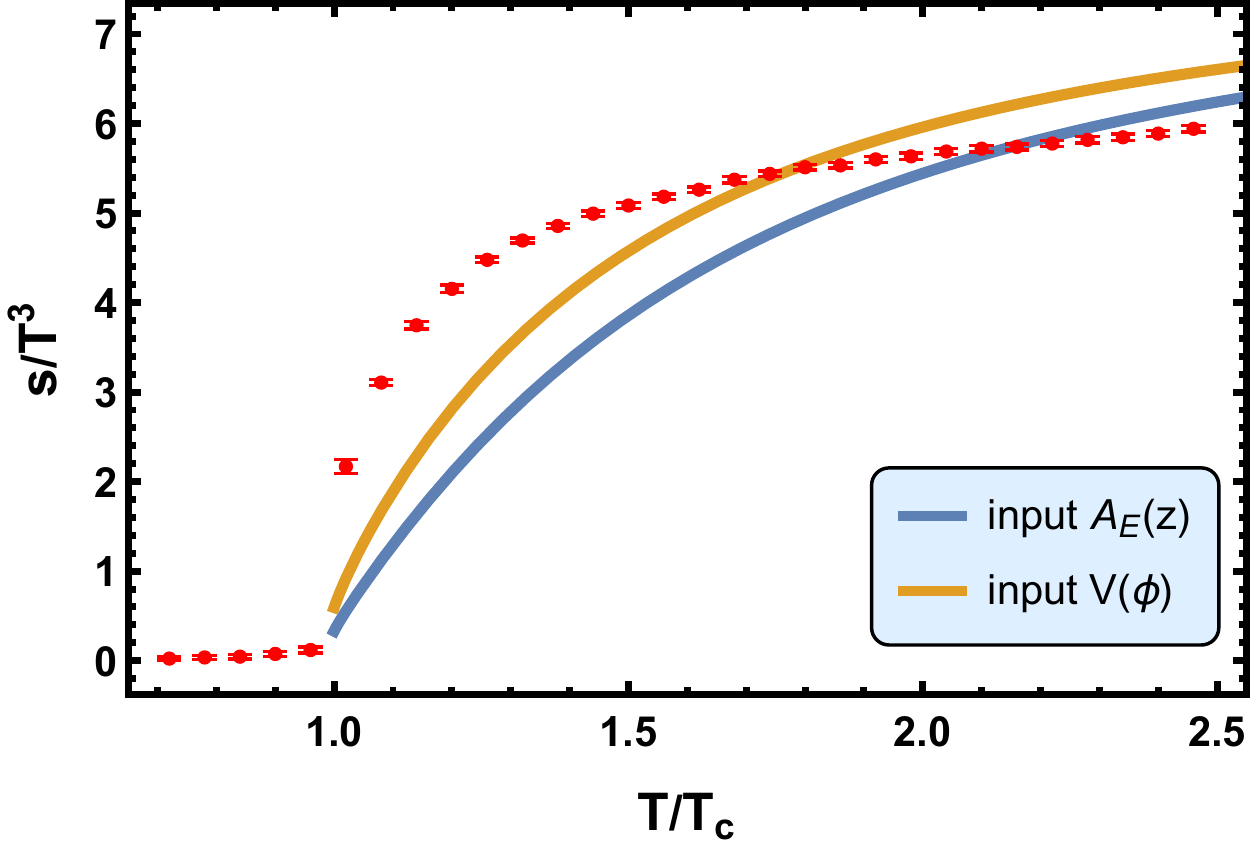}
  \hspace{0.3cm}
  \includegraphics[width=70mm,clip=true,keepaspectratio=true]{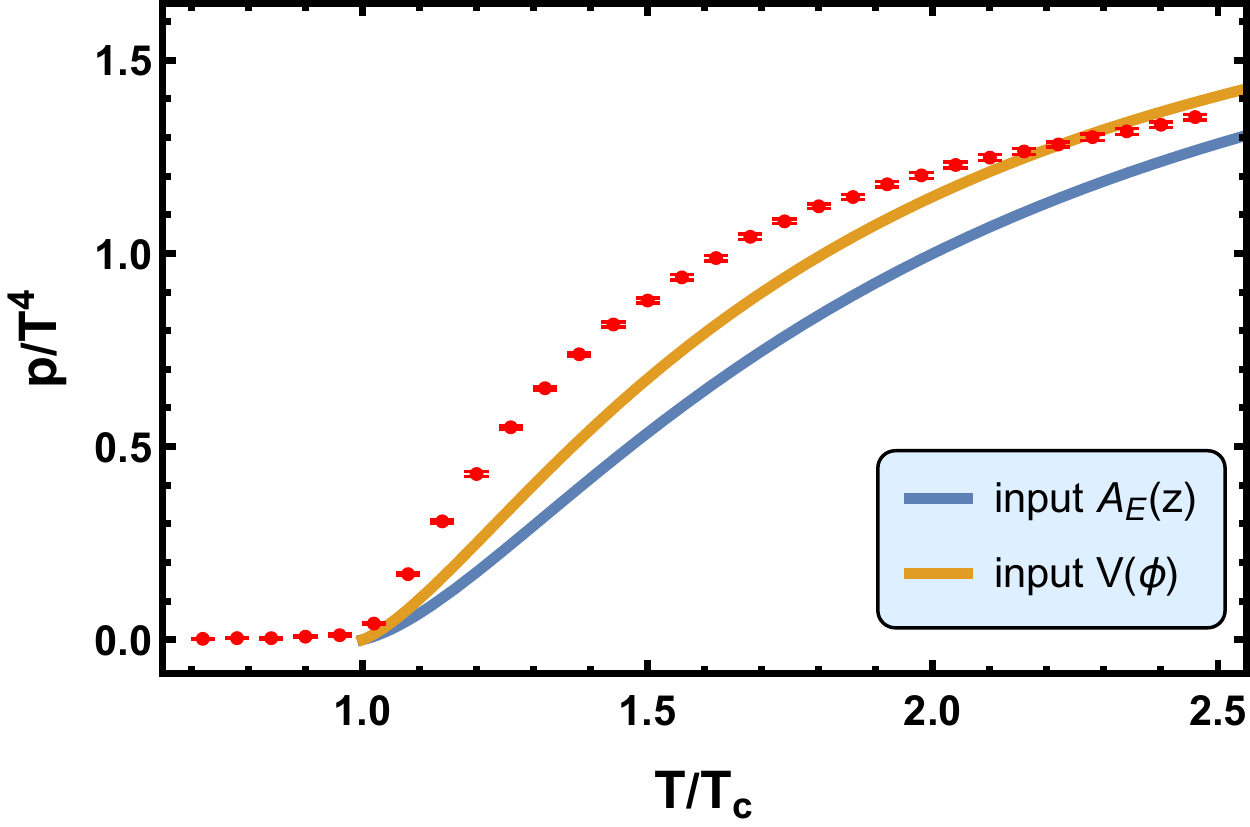}
  \vspace{0.3cm}\\
  \includegraphics[width=70mm,clip=true,keepaspectratio=true]{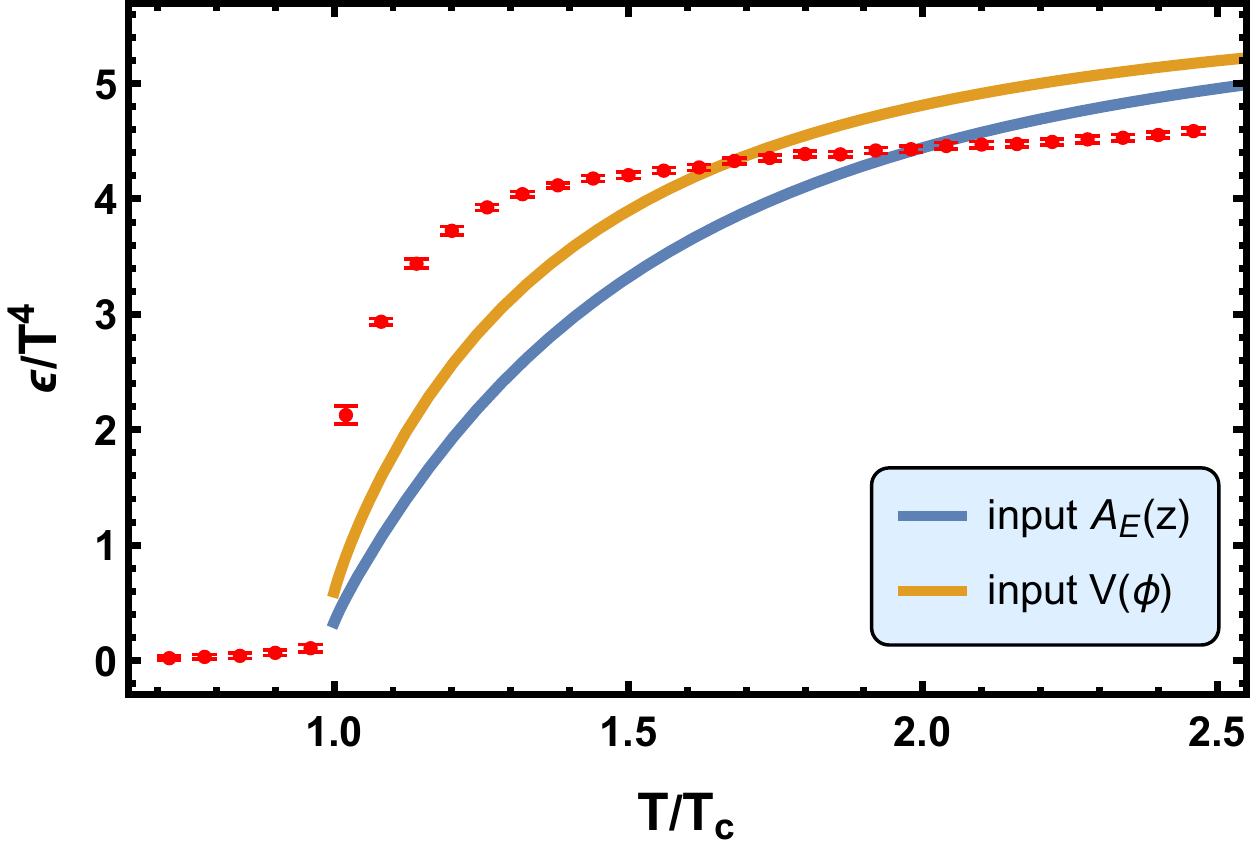}
  \hspace{0.3cm}
  \includegraphics[width=70mm,clip=true,keepaspectratio=true]{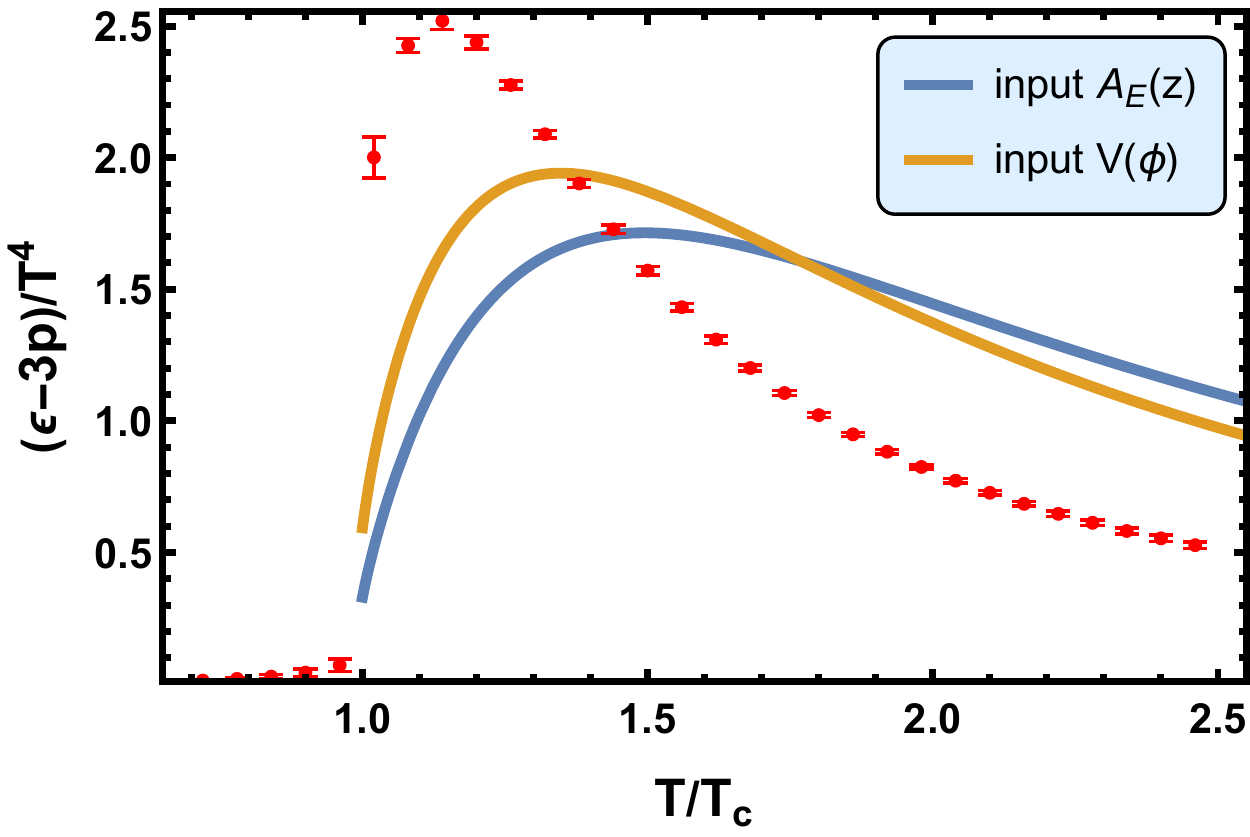}
  \caption{The results of equation of state from model \RNum{1} and model \RNum{2} with $a=0.6032 \mathrm{GeV}^2$ and $G_5=1$. The results of the entropy density over cubic temperature (upper left panel), the pressure density over quartic temperature (upper right panel), the energy density over quartic temperature (lower left panel) and the trace anomaly over quartic temperature (lower right panel) as functions of the scaled temperature $T/T_c$ in model \RNum{1} and model \RNum{2}, respectively.  The blue line is the  result for model \RNum{1} with inputting $A_E(z)$, the orange line is result for model \RNum{2} with inputting $V_{\phi}(\phi)$. The red points are $SU(3)$ lattice data taken from Ref. \cite{Caselle:2018kap}.}
  \label{Dudal_eos}
\end{figure}

In Model \RNum{1} and Model \RNum{2}, we choose the parameter $a=0.6032 \mathrm{GeV}^2$, the $5$-dimensional Newtown constant $G_5=1$. Then we numerically calculate the thermodynamical propertities for Model \RNum{1} and \RNum{2}. In model \RNum{2}, we utilize the numerical method in Refs. \cite{DeWolfe:2010he,Critelli:2017oub} to investigate the thermodynamical propertities. The results are different for these two models, as we emphasized in subsubsection \ref{model_2}. The deconfined temperature $T_c=537.960 \mathrm{MeV}$ for model \RNum{1} with inputting $A_E(z)$ and $T_c=521.147 \mathrm{MeV}$ for model \RNum{2} with inputting $V_{\phi}(\phi)$. We plot the thermodynamical quantities in Fig. \ref{Dudal_eos}. The red points with error bar is lattice simulation of $SU(3)$ Yang-Mills results in Ref. \cite{Caselle:2018kap}.

\begin{figure}[htb]
  \includegraphics[width=70mm,clip=true,keepaspectratio=true]{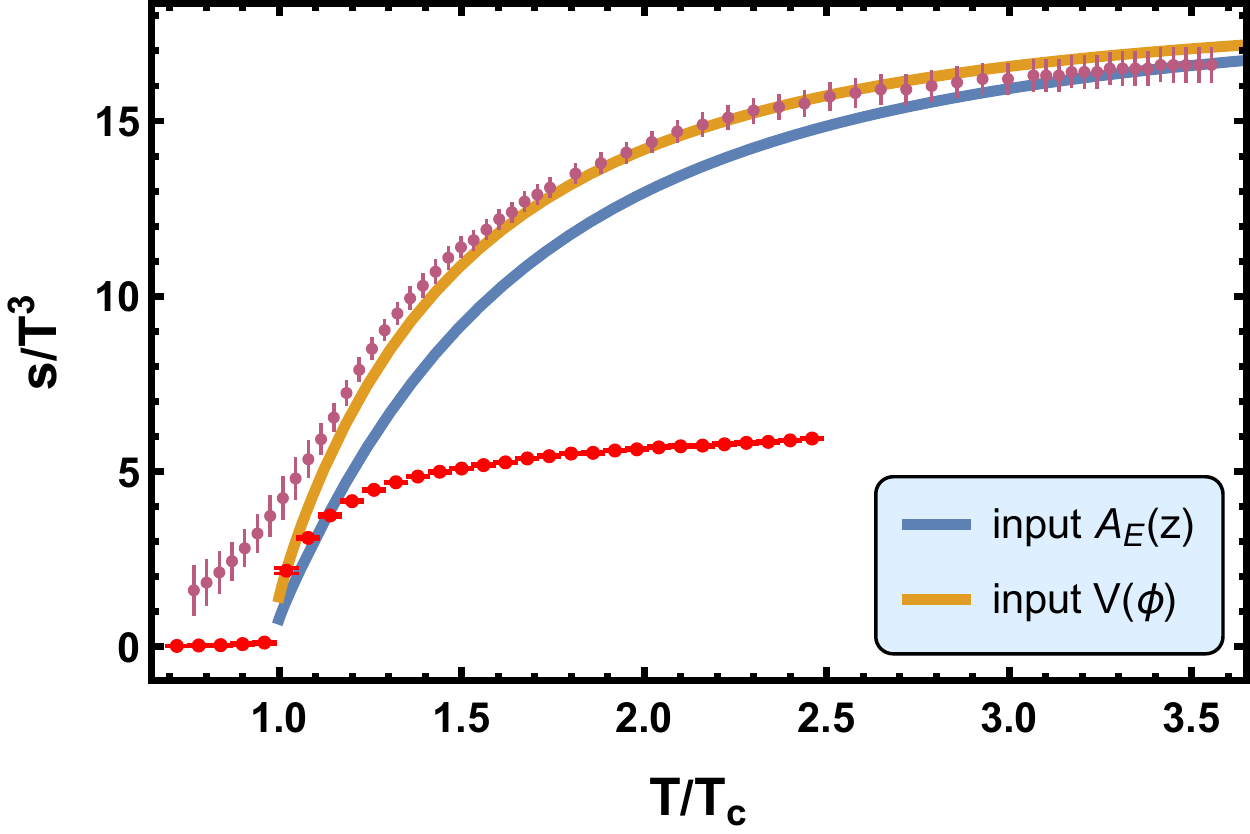}
  \hspace{0.3cm}
  \includegraphics[width=70mm,clip=true,keepaspectratio=true]{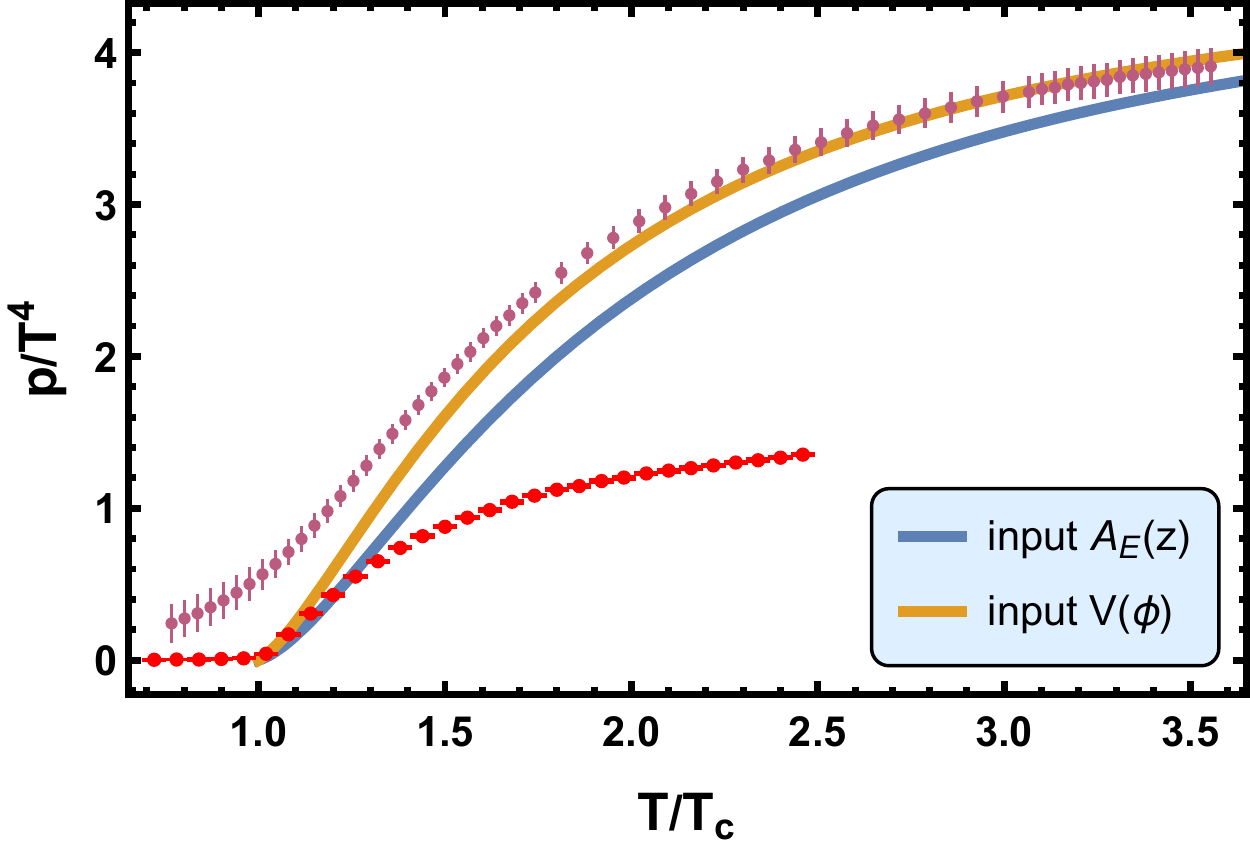}
  \vspace{0.3cm}\\
  \includegraphics[width=70mm,clip=true,keepaspectratio=true]{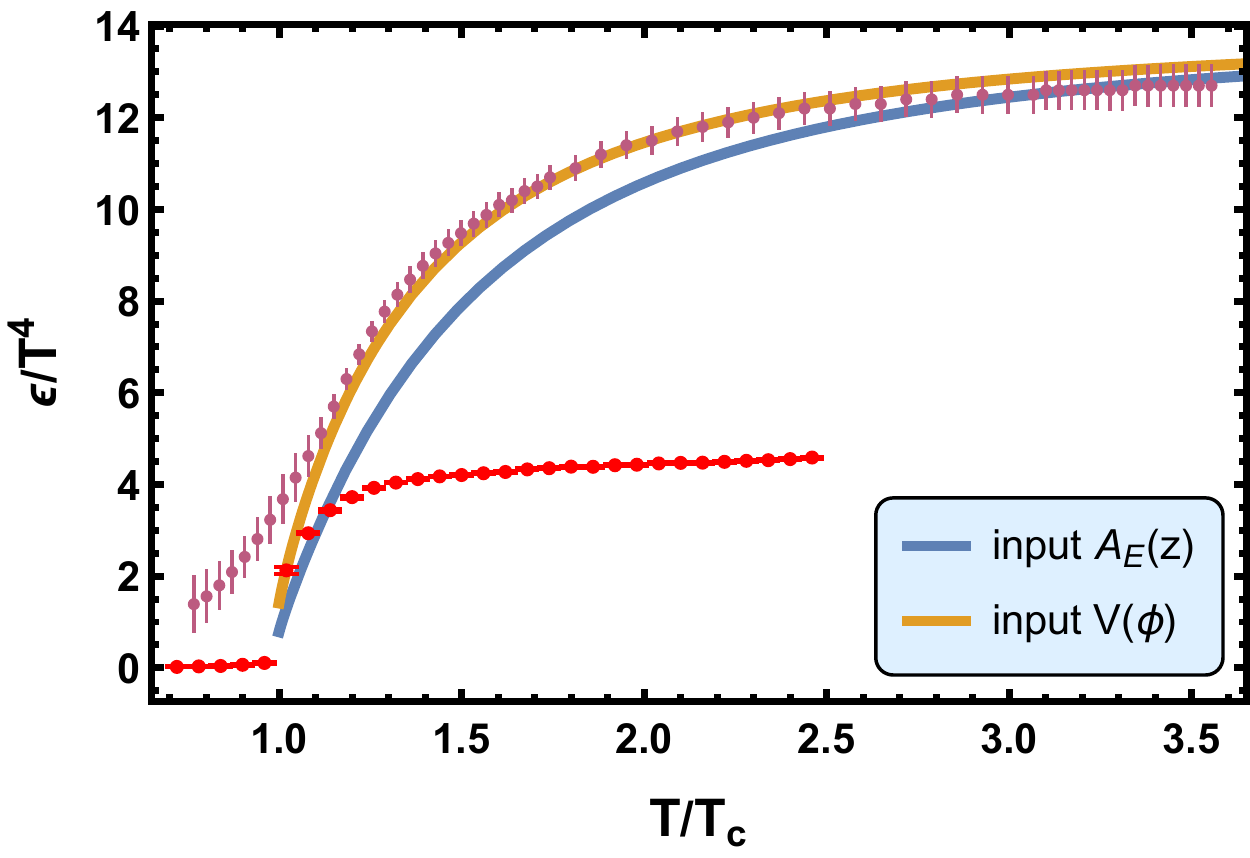}
  \hspace{0.3cm}
  \includegraphics[width=70mm,clip=true,keepaspectratio=true]{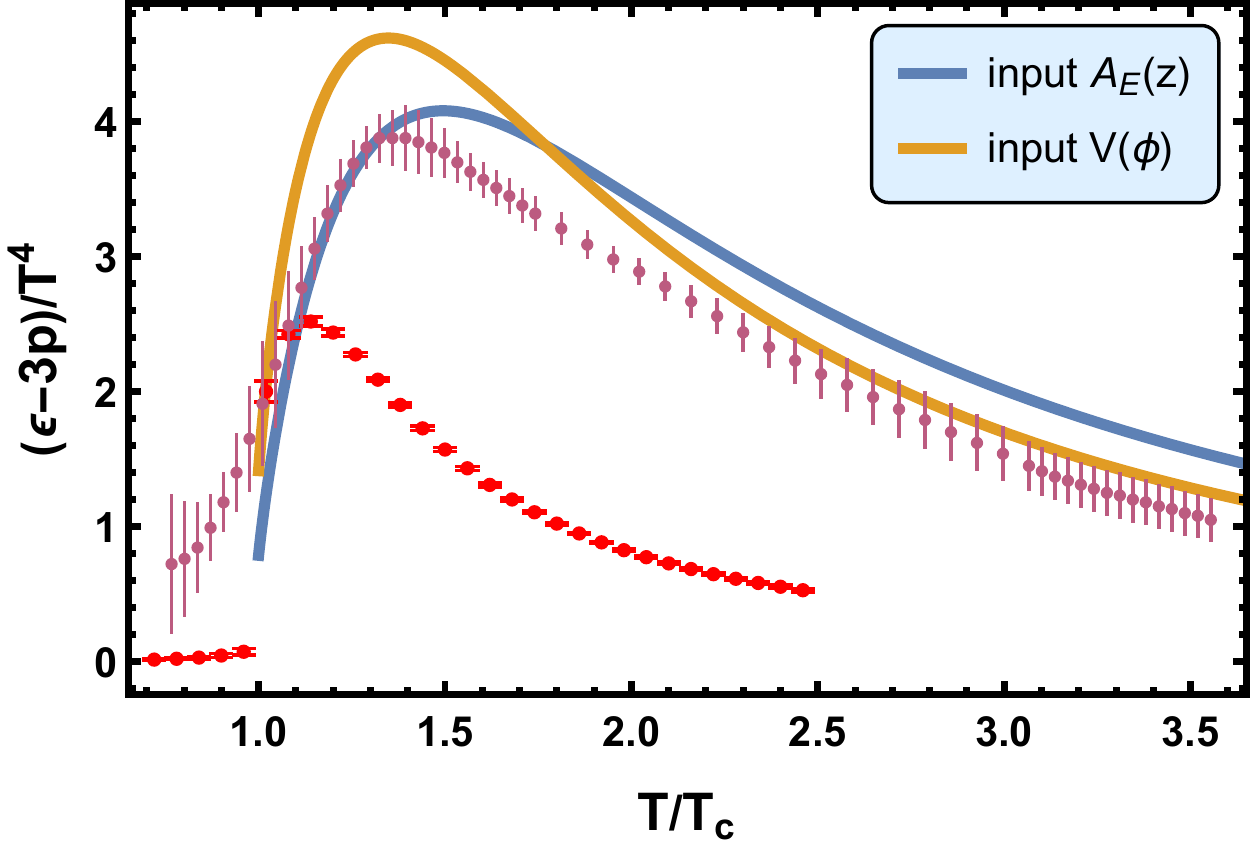}
  \caption{The results of equation of state from model \RNum{1} and model \RNum{2} with $a=0.6032 \mathrm{GeV}^2$ and $G_5=0.42$. Upper left panel: The ratio of entropy density over cubic temperature as function of scaled temperature $T/T_c$. Upper right panel: The ratio of pressure density over quartic temperature as function of scaled temperature $T/T_c$. Lower left panel: The energy density over quartic temperature as function of scaled temperature $T/T_c$. Lower right panel: The trace anomaly over quartic temperature as function of scaled temperature $T/T_c$. The blue line is for model \RNum{1} with inputting $A_E(z)$. The orange line is for model \RNum{2} with inputting $V_{\phi}(\phi)$. The red points are $SU(3)$ lattice data taken from Ref. \cite{Caselle:2018kap}, and the purple points are $N_f=2+1$ lattice data taken from Ref. \cite{Borsanyi:2013bia}.}
  \label{Dudal_eos_para_2}
\end{figure}

It is noticed that even though  Model \RNum{1} and Model \RNum{2} can describe glueballs/oddballs spectra, the corresponding thermodynamical properties shown in Fig. \ref{Dudal_eos} are not in good agreement with lattice results \cite{Caselle:2018kap} for the pure gluon system. From the asymptotic analysis of the dilaton field at UV boundary Eq. (\ref{UV_asym_1_2}) in subsection \ref{vac_sols_set_1}, we can see that the leading order of the $5$-dimensional dilaton field is a term proportional to $z$, and the sub-leading order is a term proportional to $z^3$. So we expect the thermodynamical properties of model \RNum{1} and \RNum{2} behaves more like quark matter. We fix the value of the parameter $a$,  and tune the value of $G_5=1$ to $G_5=0.42$ to meet the degrees of freedom of quark matter. In this case,  the critical temperatures remain unchanged. It is found that the equation of state calculated in model \RNum{1} and \RNum{2} are qualitatively consistent with the $2+1$ flavors lattice results in Ref. \cite{Borsanyi:2013bia}.  We plot the equation of state in Fig. \ref{Dudal_eos_para_2}. The red points with error bar is lattice simulation of $SU(3)$ equation of state taken from Ref. \cite{Caselle:2018kap} for pure gluon system. The purple points with error bar is lattice simulation of $N_f=2+1$ QCD equation of state taken from Ref. \cite{Borsanyi:2013bia}.

\subsection{Model \texorpdfstring{\RNum{3}}{3} and \texorpdfstring{\RNum{4}}{4}}
\begin{figure}[htb]
  \includegraphics[width=70mm,clip=true,keepaspectratio=true]{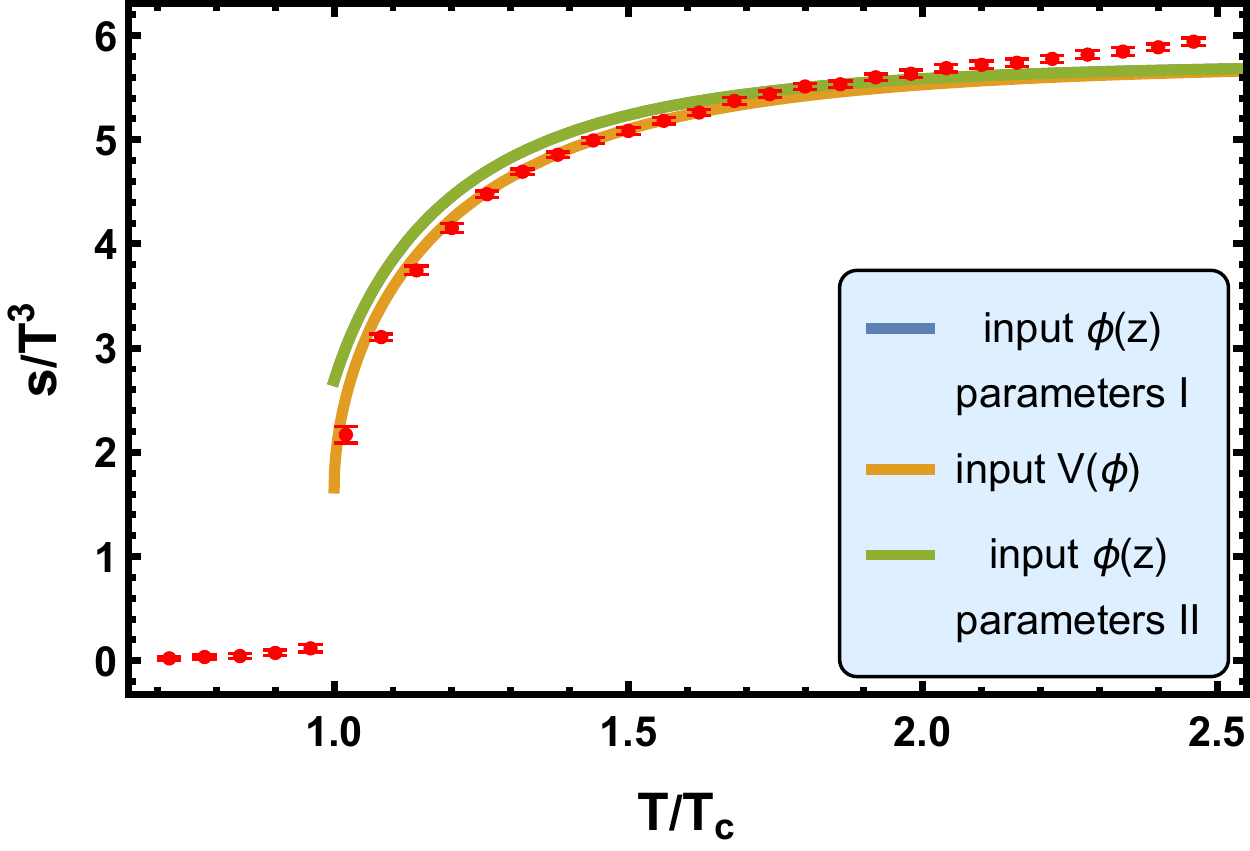}
  \hspace{0.3cm}
  \includegraphics[width=70mm,clip=true,keepaspectratio=true]{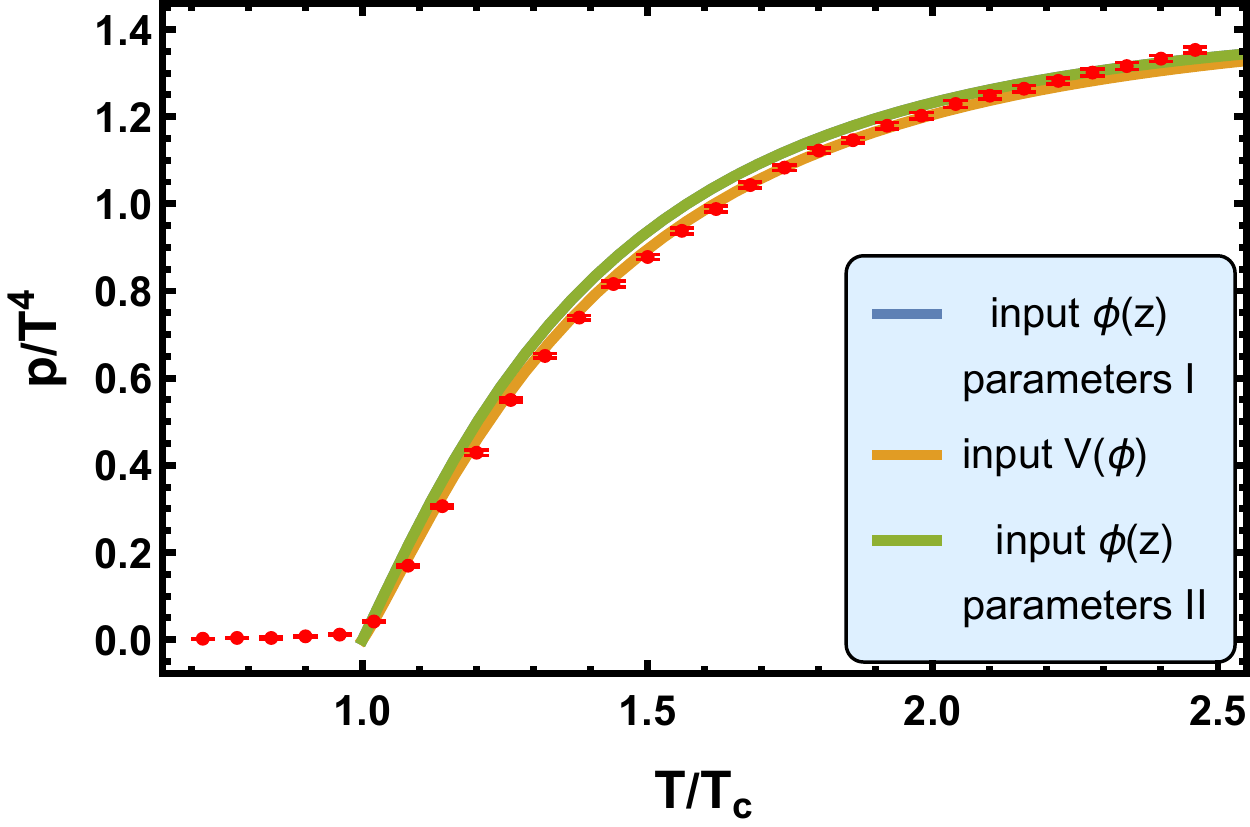}
  \vspace{0.3cm}\\
  \includegraphics[width=70mm,clip=true,keepaspectratio=true]{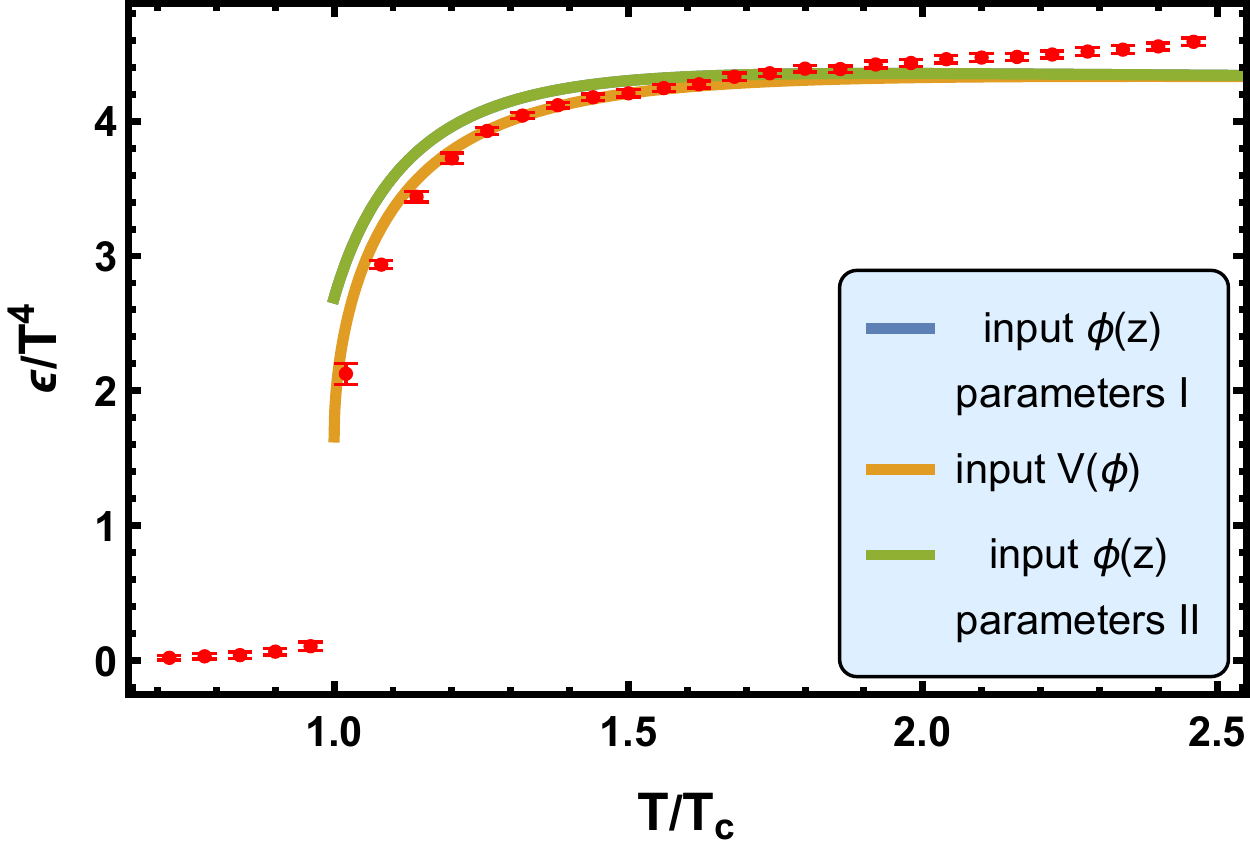}
  \hspace{0.3cm}
  \includegraphics[width=70mm,clip=true,keepaspectratio=true]{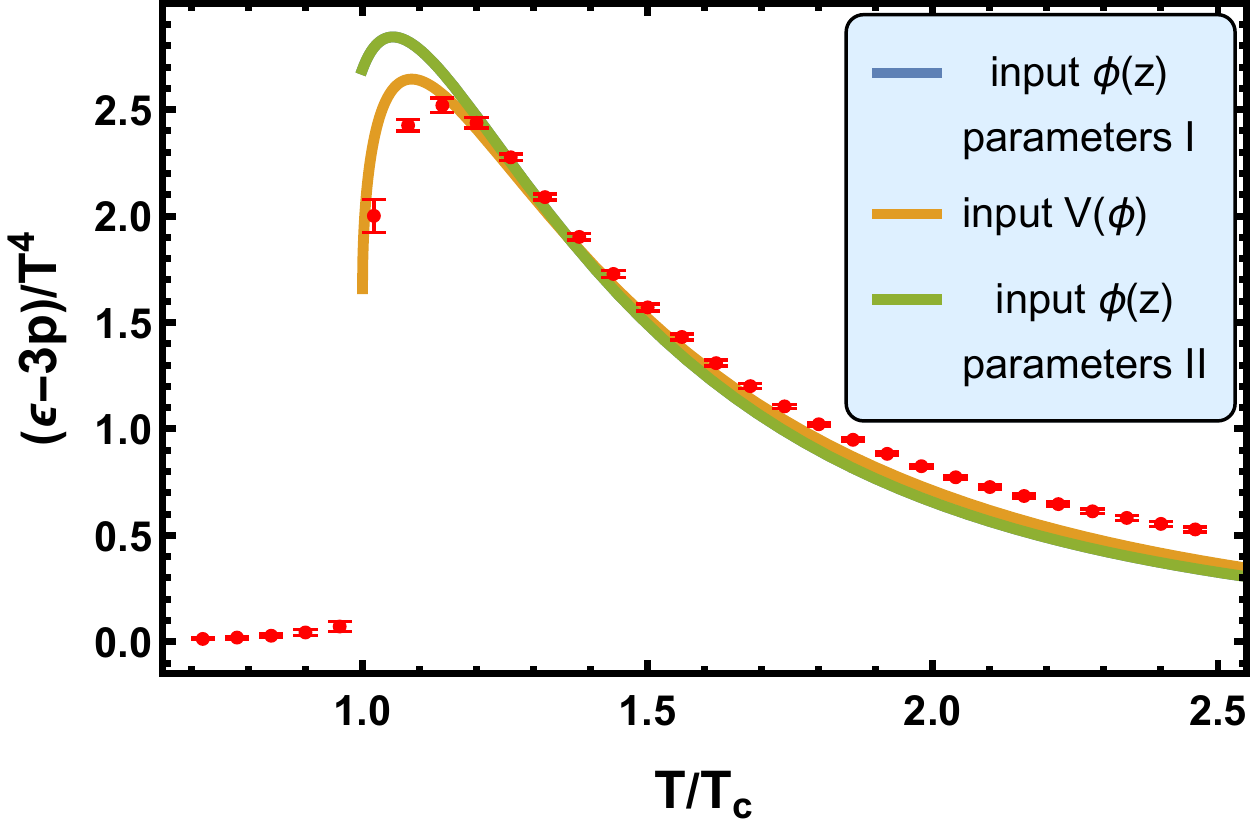}
  \caption{The results of equation of state from model \RNum{3} and model \RNum{4}. Upper left panel: The ratio of entropy density over cubic temperature as function of scaled temperature $T/T_c$. Upper right panel: The  ratio of pressure density over quartic temperature as function of scaled temperature $T/T_c$. Lower left panel: The energy density over quartic temperature as function of scaled temperature $T/T_c$. Lower right panel: The trace anomaly over over quartic temperature as function of scaled temperature $T/T_c$. The blue line is for parameters \RNum{1}: $b=1.760 \mathrm{GeV}^2$, and $G_5=1.35$ in model \RNum{3}, in which we input $\phi(z)$. The green line is for parameters \RNum{2}: $b=\frac{2 \sqrt{6}}{3} \mathrm{GeV}^2$, and $G_5=1.35$ in model \RNum{3}. The orange line is for model \RNum{4}, in which we input $V_{\phi}(\phi)$ and the parameters are $\Lambda=1 \mathrm{GeV}$, and $G_5=1.35$. The red points is $SU(3)$ lattice data taken from  Ref. \cite{Caselle:2018kap} for pure gluon sysytem. The positions of the blue line and the green line are totally the same in each panel.}
  \label{Yidian_Chen_eos}
\end{figure}

We also check the correponding thermodynamical properties of Model \RNum{3} and \RNum{4}. In model \RNum{3}, we choose two sets of parameters. The parameters \RNum{1} are $b=1.760 \mathrm{GeV}^2$, and the $5$-dimensional Newtown constant $G_5=1.35$; the parameters \RNum{2} are $b=\frac{2 \sqrt{6}}{3} \mathrm{GeV}^2$ as in \cite{Chen:2015zhh}, as we mention in subsubsection \ref{glueballs_spectra_model_3_and_4}, the $5$-dimensional Newtown constant $G_5=1.35$. Again, we employ the numerical method in Refs. \cite{DeWolfe:2010he,Critelli:2017oub} to investigate the thermodynamical propertities for model \RNum{4}. Then we fix the values of the characteristic energy scale \footnote{
  As explained in Ref. \cite{Critelli:2017oub}, the characteristic energy scale $\Lambda$, the energy dimension of which is 1, is introduced to express dimensionful observables in physical unit. When an gauge/gravity observable with dimension ${\left[E\right]}^q$ is expressed in physical unit, it should be multiplied by ${\Lambda}^{q}$.
} of the EMD system $\Lambda$ and the $5$-dimensional Newtown constant $G_5$: $\Lambda=1 \mathrm{GeV}$, and $G_5=1.35$. Then we numerically calculate the equation of state for these two models respectively. The results are actually different for the two models, as we emphasized in subsubsection \ref{model_4}. In model \RNum{3}, the deconfined temperature $T_c=367.597 \mathrm{MeV}$ for parameters \RNum{1} with $b=1.760 \mathrm{GeV}^2$ and $T_c=354.131 \mathrm{MeV}$ for parameters \RNum{2} with $b=\frac{2 \sqrt{6}}{3} \mathrm{GeV}^2$. The deconfined temperature $T_c=269.371 \mathrm{MeV}$ in model \RNum{4} with $\Lambda=1 \mathrm{GeV}$. We plot the equation of state in Fig. \ref{Yidian_Chen_eos}. The red points with error bar is lattice simulation of $SU(3)$ equation of state for pure gluon system in Ref. \cite{Caselle:2018kap}.

We can see from the Fig. \ref{Yidian_Chen_eos} that the lines from parameters \RNum{1} and parameters \RNum{2} in model \RNum{3} are totally the same with each other. That is not surprising because all the quantities are dimensionless in this plot.

\subsection{Model \texorpdfstring{\RNum{5}}{5}}
\begin{figure}[htb]
  \includegraphics[width=70mm,clip=true,keepaspectratio=true]{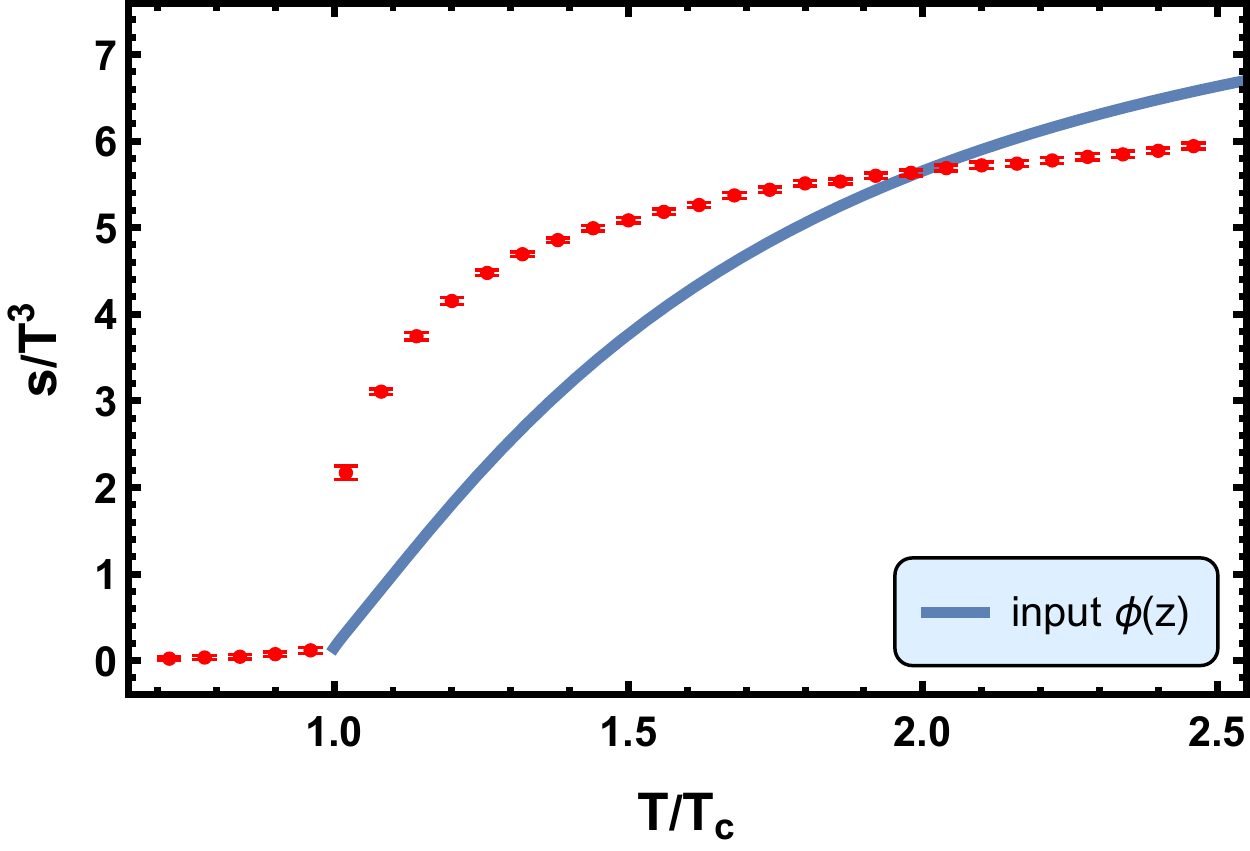}
  \hspace{0.3cm}
  \includegraphics[width=70mm,clip=true,keepaspectratio=true]{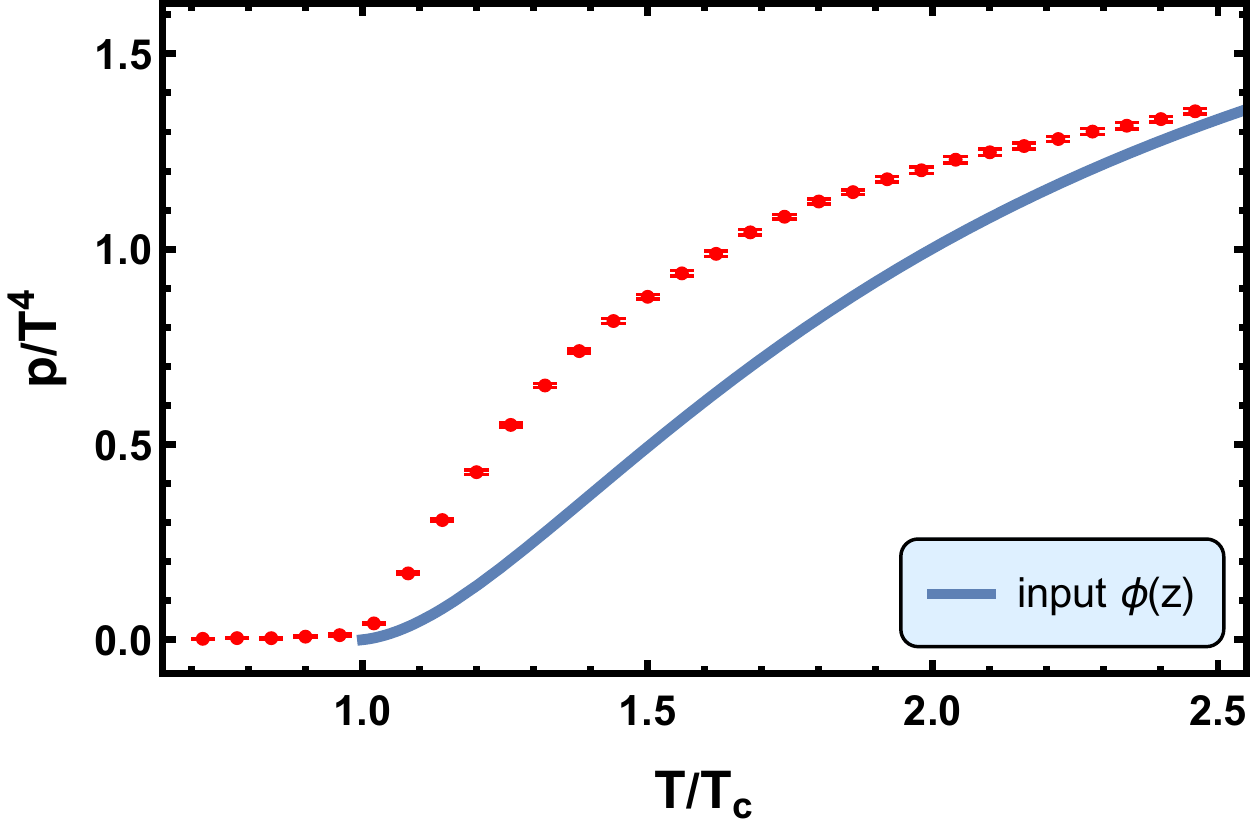}
  \vspace{0.3cm}\\
  \includegraphics[width=70mm,clip=true,keepaspectratio=true]{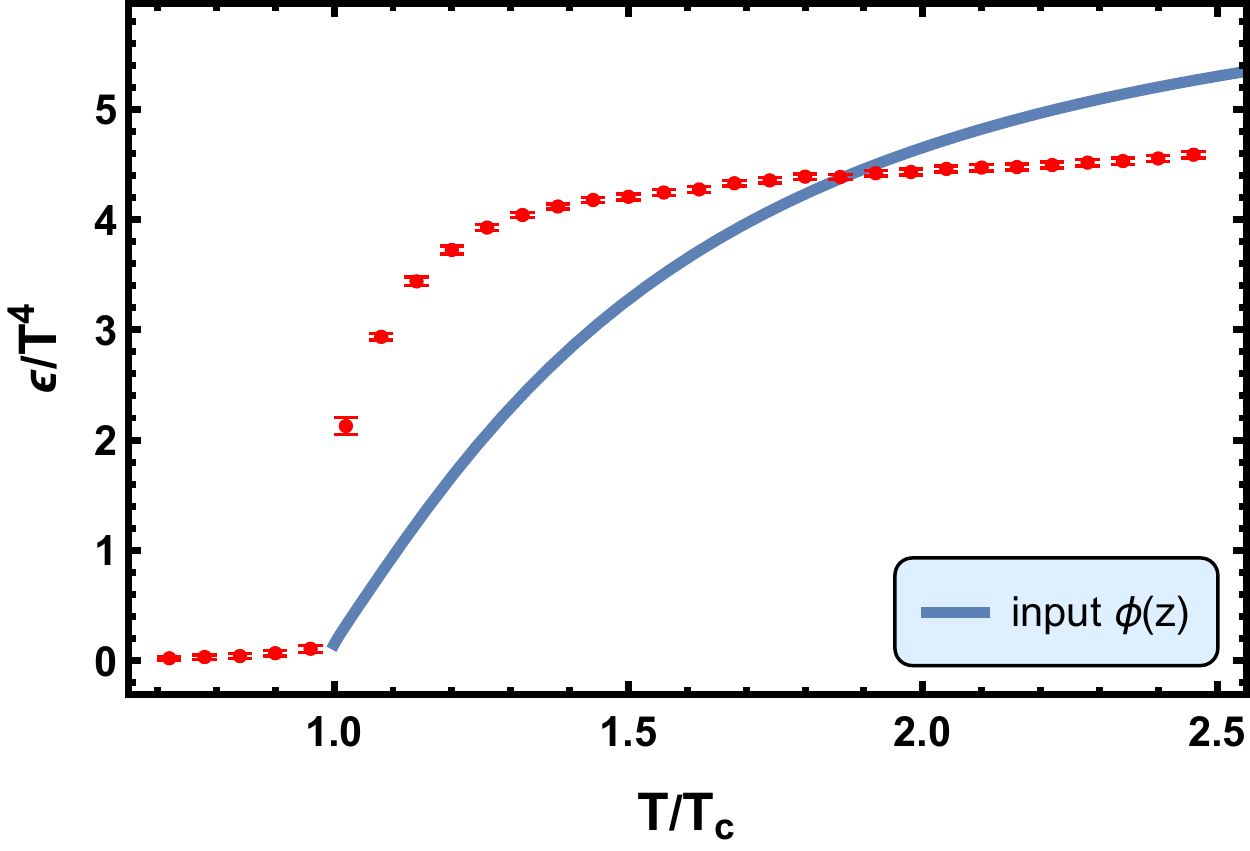}
  \hspace{0.3cm}
  \includegraphics[width=70mm,clip=true,keepaspectratio=true]{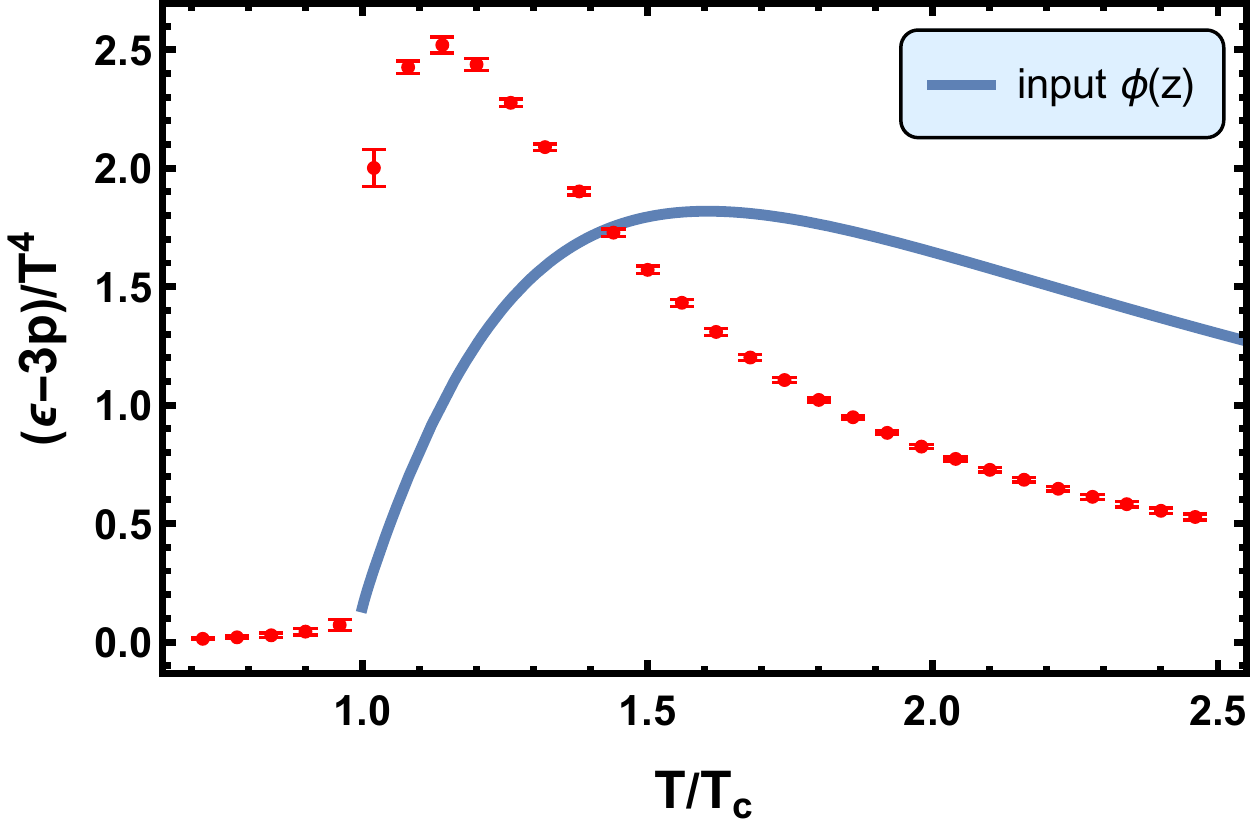}
  \caption{The results of equation of state from model \RNum{5}. Upper left panel: The ratio of entropy density over cubic temperature as function of scaled temperature $T/T_c$. Upper right panel: The  ratio of pressure density over quartic temperature as function of scaled temperature $T/T_c$. Lower left panel: The energy density over quartic temperature as function of scaled temperature $T/T_c$. Lower right panel: The trace anomaly over over quartic temperature as function of scaled temperature $T/T_c$. The blue line is for Model \RNum{5}, in which we input $\phi(z)$ and the parameters are $d=0.2 \mathrm{GeV}^2$, and $G_5=\frac{10}{11}$. The red points are $SU(3)$ lattice data taken from Ref. \cite{Caselle:2018kap} for pure gluon system.}
  \label{Chutian_Chen_eos}
\end{figure}

In Model \RNum{5}, we take the parameter $d=0.2 \mathrm{GeV}^2$, and the $5$-dimensional Newtown constant $G_5=\frac{10}{11}$. Then we numerically calculate the equation of state. The deconfined temperature $T_c=470.833 \mathrm{MeV}$. We plot the equation of state in Fig. \ref{Chutian_Chen_eos}. The red points with error bar is lattice simulation of $SU(3)$ equation of state from Ref. \cite{Caselle:2018kap}.

\begin{figure}[htb]
  \includegraphics[width=70mm,clip=true,keepaspectratio=true]{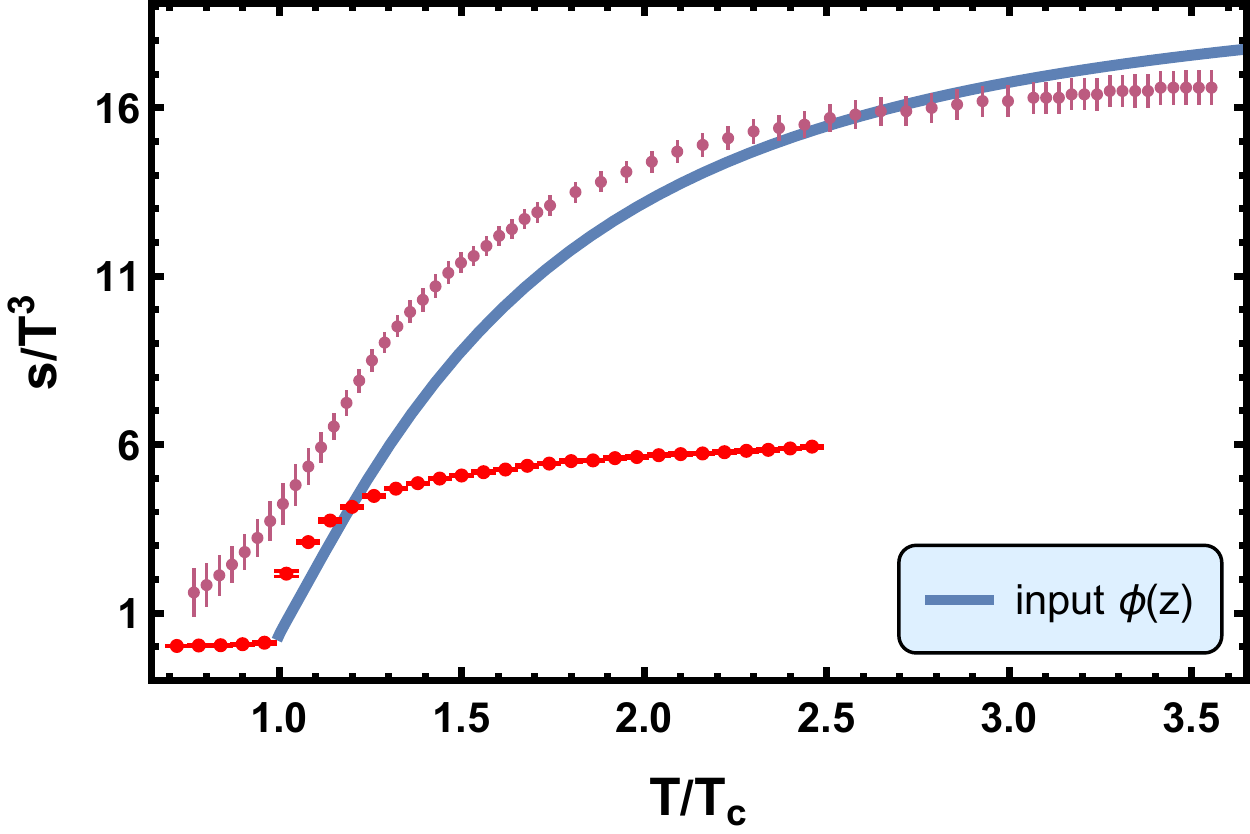}
  \hspace{0.3cm}
  \includegraphics[width=70mm,clip=true,keepaspectratio=true]{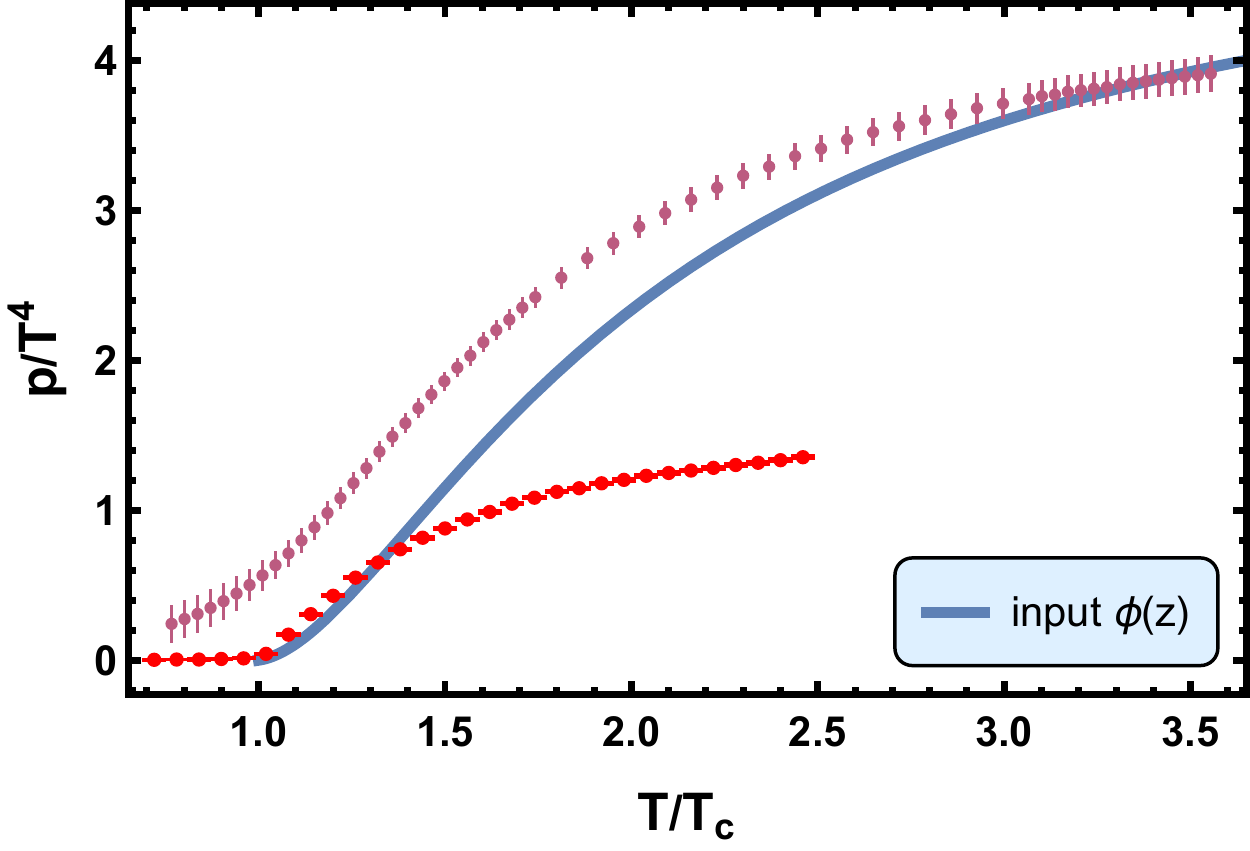}
  \vspace{0.3cm}\\
  \includegraphics[width=70mm,clip=true,keepaspectratio=true]{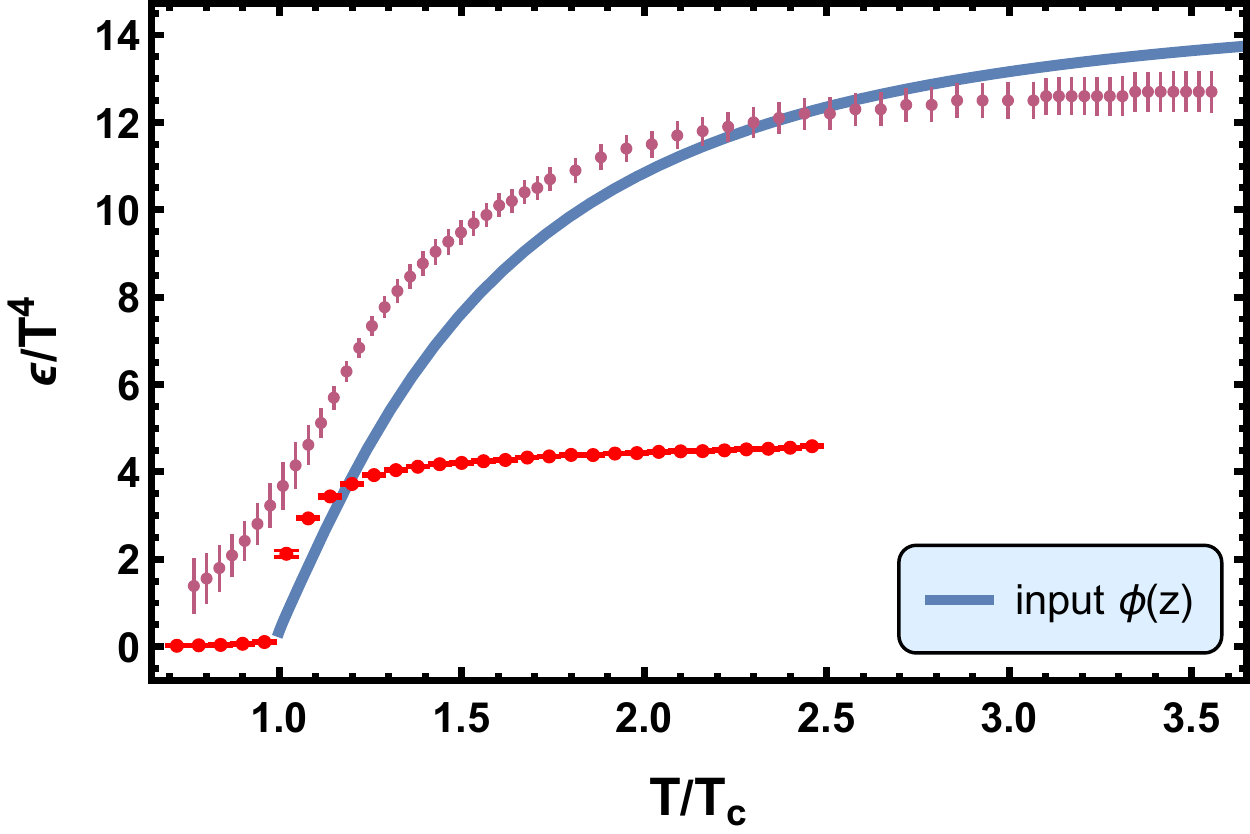}
  \hspace{0.3cm}
  \includegraphics[width=70mm,clip=true,keepaspectratio=true]{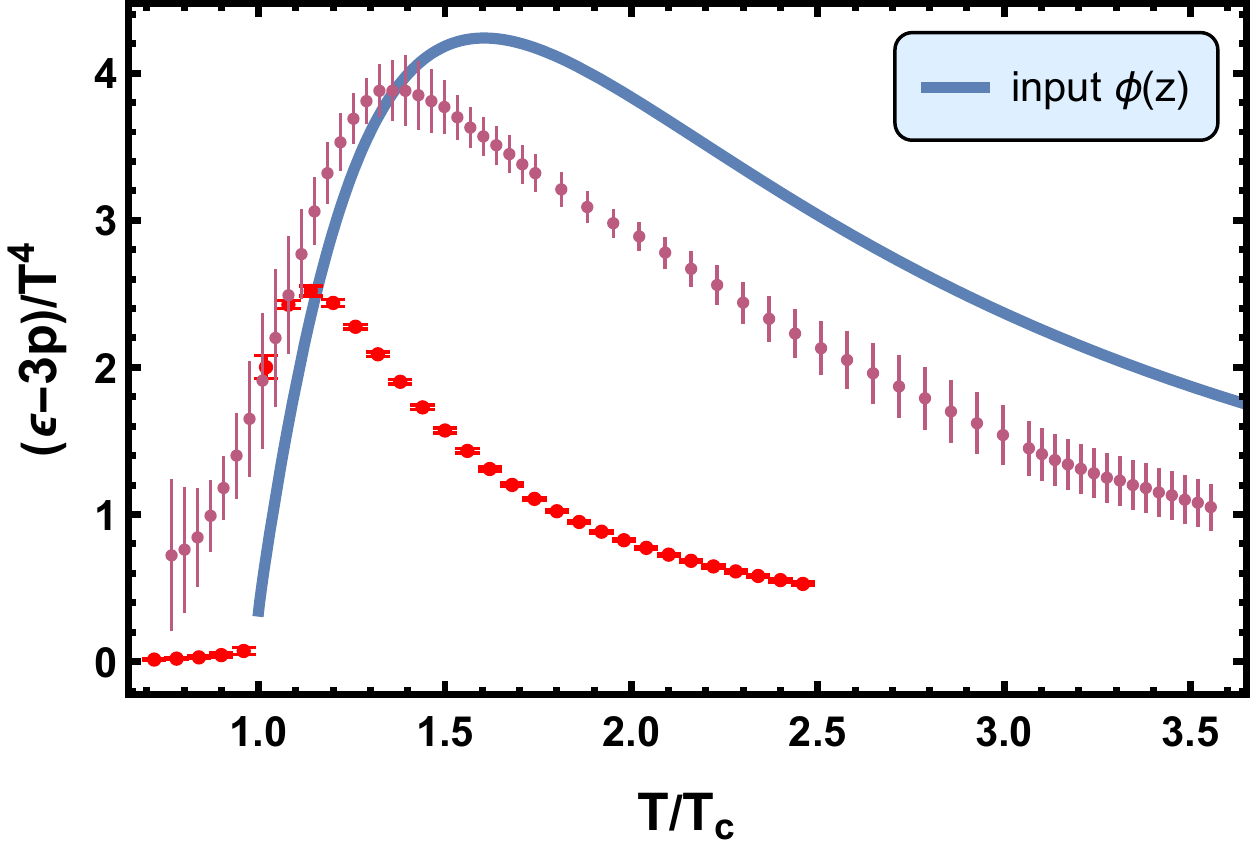}
  \caption{The results of equation of state from model \RNum{5}. Upper left panel: The ratio of entropy density over cubic temperature as function of scaled temperature $T/T_c$. Upper right panel: The  ratio of pressure density over quartic temperature as function of scaled temperature $T/T_c$. Lower left panel: The energy density over quartic temperature as function of scaled temperature $T/T_c$. Lower right panel: The trace anomaly over over quartic temperature as function of scaled temperature $T/T_c$. The blue line is for Model \RNum{5}, in which we input $\phi(z)$. The blue line is theoretical result from model \RNum{5}, in which we input $\phi(z)$ and the parameters are $d=0.2 \mathrm{GeV}^2$, and $G_5=0.39$. The red points are $SU(3)$ lattice data taken from Ref. \cite{Caselle:2018kap} for pure gluon system, and the purple points are $N_f=2+1$ lattice data taken from Ref. \cite{Borsanyi:2013bia}.}
  \label{Chutian_Chen_eos_para_2}
\end{figure}

If we fix the value of the parameter $d$ and tune the value of $G_5$ to $G_5=0.39$, the critical temperature remains unchanged. However, the equation of state in Model V will be qualitatively consistent with the $2+1$ flavors lattice results, which is taken from Ref. \cite{Borsanyi:2013bia}. We plot the equation of state in Fig. \ref{Chutian_Chen_eos_para_2}. The red points with error bar is lattice simulation of $SU(3)$ equation of state taken from Ref. \cite{Caselle:2018kap}. The purple points with error bar is lattice simulation of $N_f=2+1$ QCD equation of state taken from Ref. \cite{Borsanyi:2013bia}.

\section{Conclusion and discussion}
\label{sec-sum}

In this work, we study scalar, vector and tensor glueballs/oddballs spectra in the framework of $5$-dimensional dynamical holographic QCD model, where the metric structure is deformed self-consistently by the dilaton field. In the framework of holography, the states $J^{PC}$ with the same angular momentum $J$ and the same C\hyp{}parity share the same operator thus have the same dimension and $5$-dimensional mass, and the mass splitting for different P\hyp{}parity states is realized by $\mathrm{e}^{-p\Phi}$ in Eq. (\ref{action_glueb}). The states $J^{PC}$ with the same angular momentum $J$ and the same P\hyp{}parity but different C\hyp{}parity have different operators. Thus, they have different $5$-dimensional masses, which naturally induces the mass splitting for different C\hyp{}parity states.

From the results in Table \ref{tab_glueb_spect_holog_and_latti}, Fig. \ref{fig_glueb_spect_holog_and_latti_C_even}, and Fig. \ref{fig_glueb_spect_holog_and_latti_C_odd}, we can see that with only 2 parameters, the model predictions on glueballs/oddballs spectra in general are in good agreement with lattice results except two oddballs $0^{+-}$ and $2^{+-}$. Here we also would like to mention that the data predicted by the SP and DP Regge model \cite{Szanyi:2019kkn} to fit the high energy $pp$ scattering: using the SP Regge model, the predicted mass for $2^{++}$ glueball is $1.747 \mathrm{GeV}$; using the DP Regge model, the predicted masses for $2^{++}$ glueball and $3^{--}$ oddall are $1.758 \mathrm{GeV}$ and $3.001 \mathrm{GeV}$ respectively. These predicted values are a little bit lower than the results predicted from holography but still in reasonable regions. It might indicate that the mass $1.747 \mathrm{GeV}$/$1.758 \mathrm{GeV}$ $2^{++}$ glueball and mass $3.001 \mathrm{GeV}$ $3^{--}$ oddball are hybrid glueball/oddball states mixing with quark states.

From the results of glueballs/oddballs spectra at zero temperature and zero density and the equation of state at finite temperature, we obtain the following conclusions. 1) For the same set of vacuum solutions for the Einstein field equations and the equation of motion of the dilaton field $\phi(z)$, inputting the function $A_{E}(z)$ and inputting the dilaton potential $V_{\phi}(\phi)$ give different equation of state indeed. The difference between Model \RNum{3} and \RNum{4} are much less than that between Model \RNum{1} and \RNum{2}. 2) The model with quadratic dilaton field $\phi(z)$ can simultaneously describe glueballs/oddballs spectra as well as equation of state of pure gluon system. The model with quadratic $A_{E}(z)$ can describe glueballs/oddballs spectra, but its corresponding equation of state behaves more like $N_{f}=2+1$ quark matter. These are consistent with dimension analysis at UV boundary.

\vskip 0.5cm
  {\bf Acknowledgement}
\vskip 0.2cm

We thank Danning Li and Cong-Feng Qiao for helpful discussions.  This work is supported in part by the National Natural Science Foundation of China (NSFC)  Grant  Nos. 11735007, 11725523, and Chinese Academy of Sciences under Grant No. XDPB09, the start-up funding from University of Chinese Academy of Sciences(UCAS), and the Fundamental Research Funds for the Central Universities.
\newpage

\bibliographystyle{unsrt}
\bibliography{references.bib}

\end{document}